\documentclass[conference]{IEEEtran}

\ifCLASSINFOpdf
\else
\fi

\usepackage{cite}
\usepackage{amsmath,amsfonts,amsthm}
\usepackage{algorithmic}
\usepackage{graphicx}
\usepackage{textcomp}
\usepackage[dvipsnames, table]{xcolor}
\usepackage{xspace}
\usepackage{booktabs}
\PassOptionsToPackage{hyphens}{url}
\usepackage[hidelinks]{hyperref}
\usepackage{chronosys}
\usepackage{paralist}
\usepackage[most]{tcolorbox}
\usepackage{adjustbox}
\usepackage{multirow}
\usepackage{pifont}
\usepackage{tabularx}
\usepackage{pgfplots}
\pgfplotsset{compat=1.18}
\usepgfplotslibrary{dateplot}
\usepgfplotslibrary{statistics}
\usepgfplotslibrary{groupplots}
\usetikzlibrary{patterns}
\usepackage{wasysym}

\usepackage{framed}
\newenvironment{rqintro}%
{\begin{leftbar}\vspace{0.1em}\begin{normalsize}}%
{\end{normalsize}\vspace{0.1em}\end{leftbar}}

\usepackage{contour}
\usepackage[normalem]{ulem}

\newcommand{\myuline}[1]{%
  \uline{\phantom{#1}}%
  \llap{\contour{white}{#1}}%
}

\begin{document}
\newcommand{\etal}{et~al.\@\xspace}
\newcommand{\ie}{i.\@\,e.\@,\xspace}
\newcommand{\eg}{e.\@\,g.\@,\xspace}
\newcommand{\wrt}{w.\@\,r.\@\,t.\@\xspace}
\newcommand{\etc}{etc.\@\xspace}
\newcommand{\cf}{cf.\@\xspace}

\newcommand{\cmark}{\textcolor{PineGreen}{\ding{51}}}
\newcommand{\imark}{\textcolor{PineGreen}{\llap{(}\ding{51}\rlap{)}}}
\newcommand{\mmark}{\textcolor{orange}{\ding{51}}}
\newcommand{\emmark}{\textcolor{orange}{\ding{87}}}
\newcommand{\emark}{\textcolor{PineGreen}{\ding{87}}}
\newcommand{\xmark}{\textcolor{red}{\ding{55}}}

\tcbset{
    defstyle/.style={
        fonttitle=\upshape, 
        fontupper=\upshape,
        arc=0mm,
        colback=gray!8!white,
        colframe=black!80!gray,
        before skip=1em,        %
        after skip=1em,       %
        top=0.1em,
        bottom=0.1em,
        left=0.2em
    }
}
\newtcbtheorem%
{Insight}{Summary RQ}{defstyle}{def}

\title{Vulnerability, Where Art Thou? An Investigation of Vulnerability Management in Android Smartphone Chipsets}

\author{
\IEEEauthorblockN{Daniel Klischies}
\IEEEauthorblockA{Ruhr University Bochum\\daniel.klischies@ruhr-uni-bochum.de}
\and
\IEEEauthorblockN{Philipp Mackensen}
\IEEEauthorblockA{Ruhr University Bochum\\philipp.mackensen@ruhr-uni-bochum.de}
\and
\IEEEauthorblockN{Veelasha Moonsamy}
\IEEEauthorblockA{Ruhr University Bochum\\email@veelasha.org}
}

\IEEEoverridecommandlockouts
\makeatletter\def\@IEEEpubidpullup{6.5\baselineskip}\makeatother
\IEEEpubid{\parbox{\columnwidth}{
 		Network and Distributed System Security (NDSS) Symposium 2025\\
 		24-28 February 2025, San Diego, CA, USA\\
 		ISBN 979-8-9894372-8-3\\
 		https://dx.doi.org/10.14722/ndss.2025.241161\\
 		www.ndss-symposium.org
}
\hspace{\columnsep}\makebox[\columnwidth]{}}
\maketitle
\begin{abstract}
Vulnerabilities in Android smartphone chipsets have severe consequences, as recent real-world attacks \cite{cisaqualcomm} have demonstrated that adversaries can leverage vulnerabilities to execute arbitrary code or exfiltrate confidential information. Despite the far-reaching impact of such attacks, the lifecycle of chipset vulnerabilities has yet to be investigated, with existing papers primarily investigating vulnerabilities in the Android operating system. 
This paper provides a comprehensive and empirical study of the current state of smartphone chipset vulnerability management within the Android ecosystem. 
For the first time, we create a unified knowledge base of 3,676 chipset vulnerabilities affecting 437 chipset models from all four major chipset manufacturers, combined with 6,866 smartphone models. 
Our analysis revealed that the same vulnerabilities are often included in multiple generations of chipsets, providing novel empirical evidence that vulnerabilities are inherited through multiple chipset generations.
Furthermore, we demonstrate that the commonly accepted 90-day responsible vulnerability disclosure period is seldom adhered to. We find that a single vulnerability often affects hundreds to thousands of different smartphone models, for which update availability is, as we show, often unclear or heavily delayed. Leveraging the new insights gained from our empirical analysis, we recommend several changes that chipset manufacturers can implement to improve the security posture of their products.
At the same time, our knowledge base enables academic researchers to conduct more representative evaluations of smartphone chipsets, accurately assess the impact of vulnerabilities they discover, and identify avenues for future research.

\end{abstract}

\section{Introduction}

Smartphones play an integral part of our daily lives and are entrusted with safety-critical tasks, such as emergency calls and safeguarding of users' confidential information.
Most smartphones run a version of Android, which is the most popular mobile operating system in the world, with a market share of 70.5\% \cite{statcountermarketshare}. It is thus of utmost importance to maintain the security of Android smartphones by proactively identifying particularly vulnerable components as well as ensuring timely updates after a vulnerability has been discovered. An especially security-relevant component of smartphones is the \emph{chipset}. Chipsets consist of multiple, tightly integrated processors, some of which provide general-purpose compute to run a mobile operating system, while others provide acceleration features for graphics, or enable wireless connectivity. The functionalities implemented by chipsets inherently put them in an interesting position for an attacker, as chipsets, by design, have access to sensitive information prior to encryption as well as long-term cryptographic keys stored on the device. 

As a result of the aforementioned exposure, threat actors started to actively exploit chipset vulnerabilities. In 2023, Amnesty International reported~\cite{predator} that Predator, a commercial surveillance spyware capable of extracting messages, online browsing history, contact lists and location data from compromised phones, can be deployed through vulnerabilities in Samsung chipsets - without interaction by the victim or cooperation of the victim's service provider.
To minimize the risk of being compromised, Amnesty International recommended that individuals at-risk should ``Always update [...] as soon as any security updates are made available for your devices.'' However, this recommendation assumes the existence of device updates, which is dependent on the close cooperation of several entities, as the vulnerability originates from the chipset, rather than the device. 

Chipset processors are closely intertwined with their software, consisting of firmware, which runs on them, and the drivers that create the interface to the Android OS. Chipset processors, firmware and drivers are developed by \emph{chipset manufacturers} (CMs), rather than by smartphone manufacturers, i.e. \emph{Original Equipment Manufacturers} (OEMs), nor as a part of the \emph{Android Open Source Project} (AOSP).
This means that the CMs need to continue to support and provide updated firmware and drivers to OEMs over the lifetime of a chipset, in particular, to mitigate security vulnerabilities.
The primary goal when dealing with such vulnerabilities is to minimize the length of the \emph{vulnerability lifecycle}, i.e., the overall time frame a vulnerability is exploitable, by employing successful \emph{vulnerability management} across the supply chain. For smartphone chipsets, this time frame is divided into four phases: (i) the introduction of a vulnerability, (ii) its eventual discovery, (iii) the development of a patch removing the vulnerability by the CM, and (iv) packaging of the patch into an update that is deployed to end-users' smartphones by the OEM. Each of these phases plays a significant role in safeguarding device security, from both, a technical and an organisational perspective. Each phase has a separate technical impact, as they (i) influence which vulnerabilities a chipset will be impacted by, (ii) where discovered vulnerabilities are located, (iii) reveal how many chipset models a single vulnerability affects and how severe vulnerabilities are, and (iv) how many smartphones are ultimately affected. 

While prior work suspected that some chipset vulnerabilities tend to propagate across many generations \cite{hernandez2022firmwire, park2022doltest}, along with speculation that vulnerabilities in chipset firmware are more severe than driver vulnerabilities \cite{weinmann2012baseband}, there is a significant lack of empirical evidence to support these claims on a large scale. Existing studies have primarily relied on limited case studies or anecdotal evidence, leaving a considerable gap in our understanding of the true prevalence and impact of these vulnerabilities. Consequently, our work seeks to bridge this gap by providing extensive empirical analysis, thereby offering a more concrete foundation for future research and practical security measures. More concretely our work aims to provide a knowledge base enabling evidence-based choice of research targets, more representative evaluations and a more accurate depiction of the impact of discovered vulnerabilities.

Analogously to the technical characteristics, each phase of the vulnerability lifecycle also encompasses important organisational aspects, as they shine light on (i) the way chipset firmware and drivers, including vulnerabilities, are developed, (ii) the factors promoting the discovery of vulnerabilities, (iii) patch development timelines, and (iv) inter-organizational coordination between CMs and OEMs. These organizational aspects have recently come under scrutiny, as both, a member of Google's Threat Analysis Group\footnote{\url{https://twitter.com/maddiestone/status/1636469657136959488}} and prior work \cite{50shades} pointed out that some critical chipset vulnerabilities were not resolved within the 90 days responsible disclosure window, leaving end-users in danger after the vulnerabilities have been publicly disclosed. While such anecdotes highlight that vulnerability management processes occasionally fail, it is unclear whether this is a systemic flaw. The absence of a comprehensive, large-scale measurement undermines the ability to generalize observed process failures and suggest improvements of such processes. This is not only relevant to CMs, but to the entire security research community for two reasons: First, knowledge about internal processes should influence the decisions made by researchers when handling chipset vulnerabilities, such as the vulnerability disclosure timeline. Secondly, insights and suggestions for process improvements concerning the chipset vulnerability lifecycle may be applicable to dependent product categories where multiple companies must interact to discover vulnerabilities, develop patches and deploy these patches via device updates - such as IoT devices or connected vehicles.

\vspace{0.2em}
\noindent To determine the technical and organizational characteristics of the vulnerability lifecycle and identify potential for improvements, we devise the following research questions:

\noindent\textbf{RQ1:} Where do vulnerabilities in chipsets \emph{originate}?

\noindent\textbf{RQ2:} Who \emph{discovers} vulnerabilities in a chipset? %

\noindent\textbf{RQ3:} When are \emph{patches} available and how severe are the chipset vulnerabilities they mitigate?%

\noindent\textbf{RQ4:} What are the characteristics of the \emph{update} process utilized by OEMs to address chipset vulnerabilities in Android devices?

\vspace{0.2em}

The current lack of information on the processes that take place in every phase of the vulnerability lifecycle, preventing us from answering our RQs based on prior work, is for two reasons:
Firstly, existing papers primarily focus on understanding the final phase of the vulnerability lifecycle, in particular investigating the availability of device updates, and various factors that prolong the time it takes OEMs to provide these updates. This leaves out other phases of the vulnerability lifecycle.
Secondly, existing studies focus on vulnerabilities in, and updates for, the Android mobile operating system. The sources from which they obtain their empirical data are, however, lacking comprehensive information on chipset vulnerabilities. We make the following five main \textbf{contributions}:  %

\begin{compactitem}[-]
\item We create a unified \emph{knowledge base}, the first of its kind, aggregating comprehensive information on chipset vulnerabilities, patches, affected devices, and their update status, encompassing a majority of the Android smartphone ecosystem  (\autoref{sec:method}).
\item We then leverage our large-scale knowledge base to obtain answers to \textbf{RQ1-4}, discovering several key insights regarding the origin of vulnerabilities in newly released chipset models, limited industry transparency, and prolonged time frames for patches and updates (\autoref{sec:analysis}).
\item We compare our findings to similar studies in other ecosystems, highlighting similarities and differences. This allows us to determine factors that influence various aspects of the vulnerability lifecycle (\autoref{sec:comparison}).
\item Given our results, we discuss several actionable changes that stakeholders could implement to improve device security and enable end-users to make informed purchase decisions resulting from improved transparency (\autoref{sec:suggestions}).
\item Finally, we describe use cases on how our newly created knowledge base can enrich and enhance future research works. We elaborate how our data enables representative experimental evaluations, provides a more accurate depiction of the impact of newly discovered vulnerabilities, and unveils promising avenues for research (\autoref{sec:use-cases}).
\end{compactitem}

\noindent We \textbf{publish} our data set in the form of a continuously self-updating website at \url{https://www.chipsets.org} to enable reproducibility, allow continuous observation of future trends, and facilitate experimental evaluation.

\section{Background}\label{sec:background}

\subsection{Smartphone Chipsets}\label{sec:bg_chipsets}

Smartphone chipsets, integral components of mobile devices, are a set of specialized processors, responsible for providing various functions essential for a smartphone's operation. These chipsets encompass a diverse set of \emph{components}, including an application processor running e.g. Android, graphics processing units (GPUs), modem processors for wireless connectivity such as cellular, Bluetooth and WiFi, and various other specialized co-processors. To enable end-users to make calls, play audio over Bluetooth, or use the GPU for gaming on their smartphone, a chipset-specific driver within Android interfaces with these co-processors, delegating tasks like wireless communication or graphics computation. To this end, each co-processor executes a separate, chipset model and component-specific firmware, independently from Android. Furthermore, different components of chipsets often communicate directly with each other, \eg to synchronize on radio frequency usage \cite{classen2022attacks}. For this reason, as well as for performance and energy conservation, modern smartphone chipsets are typically manufactured in a tightly integrated way by a single CM per chipset model, rather than each functionality being supplied by a component of a different manufacturer. \cite{swanson2011greendroid, taylor2012dark, ginny2021smartphone, cabrera2021toward}.
The Android smartphone chipset market is therefore effectively an oligopoly of four CMs: Qualcomm, Mediatek, Samsung and Unisoc. These four companies cover 99\% of the smartphone chipset market \cite{counterpointsoc}. Because of this relatively small diversification, developing targeted exploits for specific chipsets is economically viable for threat actors \cite{zerodium}. This is even more of an issue, as the tight integration enables adversaries to, \eg exploit a vulnerability in a Bluetooth component to ultimately eavesdrop on WiFi communication \cite{classen2022attacks}. A holistic view on chipset vulnerabilities is thus warranted to comprehensively understand the threat-landscape and security posture of modern smartphones.

\subsection{Vantage Points}

There exists multiple vantage points from which one can obtain information on vulnerabilities and available updates.

\subsubsection*{Vulnerability Databases} The NIST National Vulnerability Database (NVD) and the CVE.org database, both operated by the US government, contain mostly identical information on vulnerabilities affecting arbitrary products, including chipsets. Due to the generic nature of such databases, they only contain a high level description of vulnerabilities, along with a Common Vulnerability Scoring System (CVSS) severeness score, which expresses the impact and exploitability of a vulnerability. The NVD also includes Common Platform Enumerations (CPEs) that, in theory, are intended to identify software and hardware affected by a vulnerability. However, not every chipset vulnerability is associated with a CPE. CPEs of chipset vulnerabilities also only specify affected chipset models, but not affected smartphone models. %

\subsubsection*{Chipset Manufacturers} Bulletins on CM websites list chipset vulnerabilities including affected model numbers, severity ratings, and usually the same description that is also later used in the NVD, as well as a CVSS score. They also contain additional, chipset-specific information, such as the component of the chipset that is affected by a particular vulnerability. CMs also acknowledge who discovered a vulnerability, if the vulnerability was not discovered by one of their employees. As each CM only publishes their own vulnerabilities, the bulletin format is not standardised across CMs.

\subsubsection*{Android Open Source Project} AOSP security bulletins provide detailed insights into vulnerabilities identified within the Android OS, their severity levels, and affected Android versions. These bulletins play a pivotal role in enhancing transparency, as they contain a unique identifier, the Security Patch Level (SPL). Each Android device also displays an SPL in its settings dialog. By matching their devices' SPL to the corresponding AOSP bulletin, end-users can assess which Android OS vulnerabilities have already received a  mitigating update, with vulnerabilities listed in bulletins not referenced by the devices' SPL being unmitigated. %
The Android Security Bulletins are therefore both: A list of vulnerabilities affecting devices, and a changelog representing which vulnerabilities have been addressed. In some cases, these bulletins also contain chipset vulnerabilities. However, AOSP security bulletins do not allow to draw an immediate conclusion on which chipset vulnerabilities a phone is vulnerable to, as this depends on the exact chipset model built into this phone. Likewise, the completeness of these bulletins \wrt chipset vulnerabilities is unclear. Therefore, only taking into account AOSP security bulletins in isolation is insufficient to obtain complete and correct information on the chipset vulnerabilities affecting a smartphone.

\subsubsection*{OEMs} In contrast to vulnerability databases and CM bulletins, OEM information is communicated on a per-smartphone, rather than per-chipset basis.
OEMs either publish device-specific changelogs on their website, listing all vulnerabilities that have been addressed in the past, or they refer to an SPL in their changelog, consequently, ignoring vulnerabilities missing from AOSP security bulletins.
OEM changelogs are thus insufficient to draw a conclusion on the security posture of a smartphone, as they only list vulnerabilities that have been already addressed, but do not enable an assessment of vulnerabilities that have not received an update.

\subsubsection*{Unifying disparate information sources} To obtain information on all chipset vulnerabilities affecting a smartphone, one currently has to (i) obtain information which chipset model is used in this smartphone, for instance from a separate data sheet, (ii) determine which vulnerabilities affect the chipset model based on AOSP, NVD, and CM information, (iii) assess the patch status of the smartphone using OEM changelogs and potentially the AOSP bulletin. Information on chipset vulnerabilities is thus scattered across multiple, disparate vantage points, necessitating a unified knowledge base in order to shed light on the chipset vulnerability lifecycle in Android smartphones.

\subsection{Vulnerability Lifecycle}\label{sec:lifecycle}

\begin{figure*}[th]
\centering
\includegraphics[width=.9\textwidth]{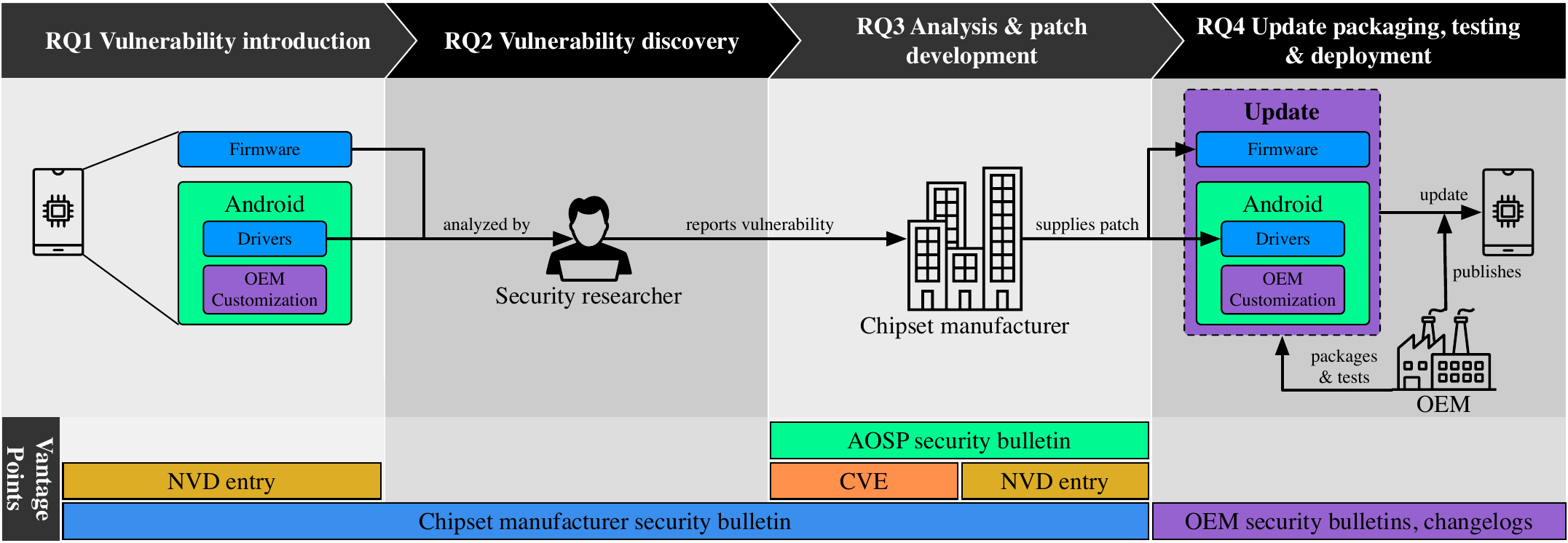}
\caption{Lifecycle of chipset vulnerabilities in the Android ecosystem.}
\label{fig:intro_timeline}
\end{figure*}

To systematically analyze the lifecycle Android chipset vulnerabilities, we adapt the linear timeline for generic software vulnerabilities, originally proposed by Schneier \cite{schneiercrypto}, to the Android chipset ecosystem as shown in Figure~\ref{fig:intro_timeline}. We identify the following four different phases, each aligning with one of our RQs:%

\subsubsection*{Vulnerability introduction (RQ1)} The lifecycle starts with a vulnerability being \emph{introduced} into a component. %
Traditionally many computer programs, such as web-browsers or Android itself, follow a continuous development model, where new features and, potentially, vulnerabilities are iteratively integrated into a single ``main'' branch of source code, from which software releases are derived \cite{firefoxdevmodel, androiddevmodel}. As a result, developers of such software typically only support and enhance the latest (few) version(s) of this branch. In contrast, chipset firmware and drivers are often specific to each chipset model \cite{hernandez2022firmwire}. %
This means that CMs might develop entirely new firmware and drivers for chipsets at different price points, and for each new chipset generation. For economical reasons, CM might also re-use and adapt parts of the source code of chipset firmware and drivers across different models. Information from NVD entries and CM security bulletins enable assessing which chipset models are affected by a vulnerability. Combining this with information on chipset release dates enables a chronological assessment of a vulnerability being introduced and potentially persisting across chipset generations.

\subsubsection*{Vulnerability discovery (RQ2)} Vulnerabilities are discovered either by the chipset manufacturers themselves, or by external third parties, such as bug-bounty hunters or academic researchers. As typical techniques for vulnerability discovery depend on the targeted component within the chipset, internal and external security experts must decide upon which component they target a-priori. This decision necessitates information on the chipset threat landscape to make an informed decision.
Information about the discovery phase is available primarily form CM bulletins, as that is the only vantage point providing information on who discovered a vulnerability as well as the affected component.

\subsubsection*{Vulnerability analysis \& patch development (RQ3)} After the discovery of a new vulnerability, CMs must analyze its potential impact and severity and develop a mitigating \emph{patch}, which they then supply to the OEMs. Finally, they publish their analysis result in the form of a CVE (that in turn is also analyzed by the NIST NVD team to assign a CVSS score), a security bulletin on their website, and, potentially, with the next AOSP security bulletin. %

\subsubsection*{Update packaging, testing \& deployment (RQ4)} 
When a vulnerability has been discovered and the OEM has been notified by the chipset manufacturer of a patch, the OEM must assess which of their phones are affected by a vulnerability, integrate driver patches into their version of Android, and package them together with firmware patches into an \emph{update} that they then deploy to the phones in the hands of their customers.
The time frame required and rigor applied in these processes is currently unknown, and thus there is currently no single source of information comprehensively summarizing for how long which phone has been exposed to a chipset vulnerability, as OEM security bulletins only contain information on mitigated, but not on unpatched chipset vulnerabilities.

\section{Related Work}\label{sec:related_work}

\begin{table}[bt]
    \centering
    \caption{Existing studies on vulnerabilities and updates in the Android smartphone ecosystem, in comparison to our study.\\ \Circle = not covered, \LEFTcircle = partially covered, \CIRCLE = covered}%
    \label{tab:related_work}
    \resizebox{1\columnwidth}{!}{%
    \small%
    \setlength\tabcolsep{1.5pt}
    \begin{tabular}{l|ccccc|cccc}
    &
        \multicolumn{5}{c}{\cellcolor{black}\color{white}\textbf{Vantage points}} &
        \multicolumn{4}{c}{\cellcolor{black}\color{white}\textbf{Phases analysed}}
        \vspace{0.2em}
        \\
        \textbf{Prior Work} &
        AOSP &
        CVE &
        NVD &
        CMs &
        OEMs &
        Intro &
        Discov. &
        Patch &
        Update
        \\
        
     \midrule
     Acar \cite{50shades} & 
     \CIRCLE & \CIRCLE & \CIRCLE & \Circle & \CIRCLE & \Circle & \Circle & \LEFTcircle & \CIRCLE
        \\
        Farhang \cite{farhang2020empirical} & 
        \CIRCLE & \CIRCLE & \Circle & \Circle & \CIRCLE & \Circle & \Circle & \LEFTcircle & \CIRCLE \\
        Hou \cite{hou2023can} &
        \Circle & \CIRCLE & \CIRCLE & \Circle & \LEFTcircle &  \LEFTcircle & \Circle & \Circle & \CIRCLE \\
        Jones \cite{jones2020deploying} &
        \LEFTcircle & \Circle & \Circle & \Circle & \LEFTcircle & \Circle & \Circle & \Circle &\CIRCLE \\
        V{\'a}squez \cite{linares2017empirical} & 
        \CIRCLE & \CIRCLE & \CIRCLE & \Circle & \Circle & \Circle & \LEFTcircle & \CIRCLE & \Circle \\
        Zhang \cite{zhang2021investigation} & \CIRCLE & \Circle & \Circle & \Circle & \LEFTcircle & \Circle & \Circle & \CIRCLE & \LEFTcircle \\
    \midrule
    Our work & 
    \CIRCLE & \CIRCLE & \CIRCLE & \CIRCLE & \CIRCLE & \CIRCLE & \CIRCLE & \CIRCLE & \CIRCLE \\
    \bottomrule
    \end{tabular}
    }
\end{table} 

\subsubsection*{Android smartphone updates}

Previous research, as shown in Table~\ref{tab:related_work}, has primarily focused on the final phase of the vulnerability lifecycle, specifically the deployment of Android OS updates ~\cite{jones2020deploying, farhang2020empirical} or device updates that include Android and chipset firmware \cite{50shades}. Hou \etal \cite{hou2023can}  focused on the final phase of the vulnerability lifecycle of Android vulnerabilities, but additionally discussed the first phase of the vulnerability lifecycle, investigating who are the developers of vulnerable, pre-installed apps. Linares-V{\'a}squez \etal focused on the second and third phase of the vulnerability lifecycle \cite{linares2017empirical}, investigating how many vulnerabilities affect each Android OS subsystem and how long it takes until vulnerabilities in the kernel are addressed. Lastly, Zhang \etal \cite{zhang2021investigation} investigated how patches propagate between Linux and different versions of Android, observing that this process is prone to delays, leaving devices exposed to kernel vulnerabilities for prolonged durations. Notably, no prior work has focused on chipset vulnerabilities or included all vantage points that are required to gain a holistic picture of chipset vulnerability lifecycle, with CMs' websites websites being a common oversight. Furthermore, the early phases of the vulnerability lifecycle have not been studied comprehensively, as existing papers mostly focus on its final phase.

\subsubsection*{Generic software} There exist studies that have investigated the lifecycle of vulnerabilities in PC software~\cite{shahzad2019large, alexopoulos2022long, arbaugh2000windows, frei2006large, alexopoulos2020tip}. Prior work that studied whether vulnerabilities persist across several versions of software have discovered that this heavily depends on the kind of software at hand; for instance, web browsers show very different characteristics compared to operating systems~\cite{clark2014moving,alexopoulos2020tip, shahzad2019large, clark2010familiarity, 7802638}. There exist few studies investigating how product security teams of software vendors fare in comparison to external security researchers, exclusively targeting Firefox and Chrome~\cite{sivagnanam2021benefits, atefi2023benefits}. When considering discoveries made by external security researchers, the effectiveness of bug bounty programs is an active research area~\cite{alexopoulos2020tip, zhao2015empirical}. However, these works typically investigate bug-bounty programs that have a low barrier of entry by not requiring physical access to devices, such as bug bounties on web applications. Results of prior works might thus not translate to programs that require access to a specific phone, such as bug bounties on chipset software. Lastly, prior works on vulnerability analysis, patch development, and update deployment processes studied the reliability of vulnerability analyses~\cite{anwar2021cleaning, dong2019towards,jiang2021evaluating}, difficulties in replicating vulnerabilities for verification~\cite{mu2018understanding, croft2022investigation}, and update timelines  \cite{atefi2023benefits, shahzad2019large}.

\section{Method}\label{sec:method}

\subsection{Data Collection}\label{sec:knowledge_base}

We collected vulnerability information from three independent vantage points: (i) chipset manufacturer security bulletins of Samsung, Qualcomm, Mediatek and Unisoc, (ii) NIST's NVD, and (iii) AOSP security bulletins. This allows us to later uncover any inconsistencies, if present, between the three sources when investigating our research questions. We only consider vulnerabilities of those chipset models that include at least the two basic components required to build a cellular phone, \ie an application processor, running Android, and a cellular modem. Vulnerabilities in other components of such chipset models are also in scope. We list all components in \autoref{appendix:location_component}. This results in information on 3,676 different vulnerabilities across 437 chipset models, released between September 2009 and April 2024.
\emph{Our dataset thus includes chipset and vulnerability information on all relevant CMs, resulting in a market coverage of 99\% \cite{counterpointsoc}.}

We aim to cover a diverse set of OEMs, while ensuring that our analysis results are not heavily skewed by legacy information. To avoid introducing a negative bias from defunct OEMs who have stopped updating their devices, we only included OEMs who have released a new smartphone model since the beginning of 2022. Querying GSMArena\footnote{\url{https://GSMArena.com}}, a website that independently curates information on smartphones, based on this criterion resulted in smartphone to chipset mappings and release date information covering 6,866 smartphone models made by 38 OEMs.
Notably, only four out of 38 OEMs (Samsung, Google Pixel, Tecno, and Fairphone) provide comprehensive changelogs of all their devices on their websites, with a third-party website offering data for Xiaomi. Three of these OEMs rank among the top four in Q2 2023 Android smartphone sales \cite{idcsmartphone}, while Google Pixel and Fairphone have negligible market shares \cite{statistapixel, fairphone}.
We therefore also include information on 16,139 device updates - covering individual device updates as well as information from OEM Security Response Centers - for all smartphone models manufactured by Samsung, Tecno, and Xiaomi into our dataset.
\emph{Our data thus covers the majority of the Android market regarding information on which chipset is used in which phone, as well as device update information \cite{idcsmartphone}.} 
Additional technical details of this process are described in \autoref{appendix:jsonhtml}, with \autoref{tab:vantage_point_urls} providing a summary of vantage points.

\subsection{Data Augmentation}\label{sec:augmentation}

Since the scope of our RQs requires information that goes beyond the data immediately contained within the vantage points, we augment our dataset by inferring additional information before aggregating all collected information into a%

\subsubsection*{Affected Component}\label{augmentation:component_location} Chipset manufacturers do not provide information on which component was affected by a vulnerability that uses consistent names throughout different manufacturers. Instead, CMs often refer to components by internal marketing or code names. For instance, the Trusted Execution Environment (TEE) used in Mediatek chipsets is called 'Kinibi', while Qualcomm refers to their TEE as 'QSEE'. For this reason, we identified a set of such key terms for each component based on source code fragments, data sheets, and marketing material published by each CM. We then use these key terms to automatically identify a vulnerability's component based on information provided in CM security bulletins, normalizing to a common component name that we use across all CMs (\eg 'Trust' in the example above), according to the list in Appendix \autoref{appendix:location_component}.

\subsubsection*{Vulnerability Location}
The \emph{location} specifies whether the vulnerability is located within a driver executing as a part of Android, or if it the vulnerable code is contained in the component's firmware, executing on a separate co-processor. We identify a vulnerability's location using the same approach as we use for components.%

\subsubsection*{External reports}\label{sec:04_external_reports} Only Qualcomm explicitly distinguishes between external and internal sources in their security bulletins, while Mediatek, Unisoc, and Samsung credit external discoverers, but not their own employees, by name for each vulnerability if the researcher consents to this. Vulnerabilities with named discoverers are, therefore, always  external discoveries. Depending on how many external reporters want to remain anonymous, the true number of externally found vulnerabilities might be higher than the estimated heuristic for Samsung, Mediatek, and Unisoc. As we will see in Section~\ref{sec:who_discovers}, this possible under-approximation does not invalidate but strengthens our empirical results.

\subsubsection*{Knowledge Base}
Afterwards, we build our knowledge base by combining information from various vantage points to fit the canonical schema illustrated in  Figure~\ref{fig:method_information_relations} (Appendix \autoref{appendix:integration}). To do so, all collected and augmented data is stored in a relational database. We link individual data points using common identifiers, such as CVE IDs for vulnerabilities or model numbers for chipsets. This allows us to correlate and compare information obtained from different vantage points.

\section{Empirical Analysis}\label{sec:analysis}

\subsection{RQ1: Vulnerability Introduction}\label{sec:rq1}\label{sec:rq1_origin_of_vulns}

\begin{rqintro}
\noindent\emph{Where do vulnerabilities in chipsets originate?}
\end{rqintro}
\begin{figure}[th]
\centering
\vspace{-1em}
\includegraphics[width=.95\columnwidth]{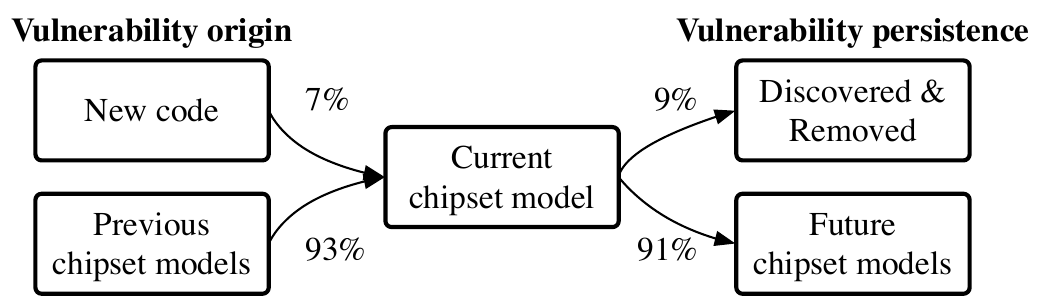}
\caption{Origin and persistence of discovered vulnerabilities in the average chipset model. Arrows symbolize the direction in which vulnerabilities propagate.}
\label{fig:result_intro_sankey}
\end{figure}

\subsubsection{Vulnerability origin} To assess if vulnerabilities introduced through \emph{code re-use} from previous chipset generations outweigh newly introduced ones, we measure the amount of vulnerabilities that have been newly introduced into every chipset, and compare this to the amount of all vulnerabilities affecting each chipset. To do so, we utilize the information about which chipsets are affected by a vulnerability, as published in CM security bulletins and NVD entries, combined with the release date of each chipset. We associate each chipset $c \in C$ with a set of vulnerabilities $V(c)$ that affected the chipset, and the chipset's release date $T_\text{rel}(c)$. Then a vulnerability $v$ has been \emph{newly introduced} in chipset $c$ iff. $\forall c' \in C . (v \in V(c') \rightarrow T_\text{rel}(c') \geq T_\text{rel}(c))$.

We observe that each chipset $c$ is affected by $|V(c)| = 204$ vulnerabilities on average, with a median of 149. However, on average, only 7\% of the vulnerabilities affecting a chipset have been newly introduced in a particular generation of chipset models, while 93\% of vulnerabilities are inherited from previous generations, as shown on the left side of Figure~\ref{fig:result_intro_sankey}. In the median case, a new chipset model does not introduce any new vulnerabilities. Thus, the distribution is heavily skewed, and there are a few chipset models that, upon release, introduce the majority of new vulnerabilities. The primary origin for vulnerabilities in newly released chipset models are thus vulnerabilities inherited from previous chipset generations through code re-use, rather than new vulnerabilities, which only occur in very specific chipset model releases.

We also found that the introduction of support for new cellular, WiFi or Bluetooth protocols does not always lead to the highest percentage of newly introduced vulnerabilities. Instead, we observe that chipset models supporting identical protocols have vastly different shares of newly introduced vulnerabilities. For instance, we discovered that Qualcomm's SM8475 chipset exhibited 21\% newly introduced vulnerabilities, surpassing the 6\% found in its predecessor, the SM8450, despite both supporting the same cellular, WiFi and Bluetooth protocols. In comparison, Qualcomm's first 5G supporting chipset models SM4350 (7.5\%), SM8150 (8\%) and SM8350 (13.5\%) also exhibit a smaller share of newly introduced vulnerabilities than the SM8475. The same holds for other CMs: 18\% of the vulnerabilities in Mediatek's MT6889, released in the first quarter of 2020 and their first model to support 5G, were newly introduced. The MT6762, only supporting LTE and released in 2018, a time when LTE was already widely adopted, saw a comparable rate of 17.5\% of new vulnerabilities. Therefore, we suspect that, while the introduction of 5G led to above-average rates of newly introduced vulnerabilities, similar percentages of newly introduced vulnerabilities can be likewise caused by internal changes made by the CM. 

\subsubsection{Vulnerability persistence} We determine whether vulnerabilities \emph{persist} from one chipset generation to the next. Formally, if $T_\text{patch}(v)$ is the date at which vulnerability $v$ has been patched, we consider $v$ to persist into chipset $c$ iff $v$ was not newly introduced in $c$ and $T_\text{rel}(c) \leq T_\text{patch}(v) \wedge v \in V(c)$ holds.
We find that, on average, only 9\% of vulnerabilities of a chipset model are removed before the release of the next chipset model by the same CM ($T_\text{rel}(c) \geq T_\text{patch}(v)$), while 91\% of vulnerabilities persist ($T_\text{rel}(c) < T_\text{patch}(v)$), as shown on the right side of Figure~\ref{fig:result_intro_sankey}. The distribution is heavily skewed, as in the median case no vulnerabilities are removed. Combined with our previous results, we observe a decreasing trend in the amount of vulnerabilities in chipsets: On average, each new chipset generation leads to the addition of 6\% of new vulnerabilities, while 9\% of the vulnerabilities are removed.
\begin{Insight}{Vulnerability introduction}{placeholder}
\myuline{\textbf{Observations:}}
\begin{compactitem}[$\circ$]
    \item For new chipset models, 93\% of vulnerabilities are inherited from previous chipset generations, rather than  newly introduced by novel features.
    \item On average, new chipset models tend to exhibit less vulnerabilities than the previous generation, as there are fewer vulnerabilities introduced than discovered.
\end{compactitem}
\myuline{\textbf{New insight:}}
\begin{compactitem}[$\circ$]
    \item Vulnerabilities affecting cutting-edge chipset models may be found by testing legacy devices, demonstrating that it is often-times unnecessary to acquire new devices to find practically relevant vulnerabilities. 
\end{compactitem}
\end{Insight}

\subsection{RQ2: Vulnerability Discovery}\label{sec:rq2}

\begin{rqintro}
\noindent\emph{Who discovers vulnerabilities in a chipset?}
\end{rqintro}

\begin{figure*}[ht]
\centering
\begin{tikzpicture}

\begin{axis}[
    height=13em,
    width=0.90\textwidth,
    ybar,
	bar width=6pt,
    legend style={at={(1,1)},
      anchor=north east,legend columns=1},
    ylabel={\# Vulnerabilities},
    ylabel style={align=center, font=\small},
    date coordinates in=x,
    xtick distance=365.5,
    reverse legend,
    axis y line*=right,
    axis x line=none,
    date ZERO=2015-01-01
    xmin=2015-01-01
    ]
\addplot+[ybar, preaction={fill, blue!30!white}, draw=blue, text=black] plot coordinates {
(2015-01-01, 1)
(2016-01-01, 6)
(2017-01-01, 9)
(2018-01-01, 15)
(2019-01-01, 23)
(2020-01-01, 48)
(2021-01-01, 38)
(2022-01-01, 16)
(2023-01-01, 43)
};

\addplot+[ybar, preaction={fill, red!30!white}, draw=red, text=black] plot coordinates {
(2015-01-01, 0)
(2016-01-01, 0)
(2017-01-01, 0)
(2018-01-01, 295)
(2019-01-01, 272)
(2020-01-01, 344)
(2021-01-01, 268)
(2022-01-01, 220)
(2023-01-01, 314)
};

\addplot+[ybar, fill=green!30!white, draw=green, text=black] plot coordinates {
(2015-01-01, 0)
(2016-01-01, 0)
(2017-01-01, 0)
(2018-01-01, 0)
(2019-01-01, 0)
(2020-01-01, 1)
(2021-01-01, 2)
(2022-01-01, 104)
(2023-01-01, 465)
};

\addplot+[ybar, fill=orange!30!white, draw=orange, text=black] plot coordinates {
(2015-01-01, 0)
(2016-01-01, 0)
(2017-01-01, 0)
(2018-01-01, 0)
(2019-01-01, 1)
(2020-01-01, 1)
(2021-01-01, 95)
(2022-01-01, 265)
(2023-01-01, 317)
};

\end{axis}

\begin{axis}[
width=0.9\textwidth,
height=13em,
date coordinates in=x,
axis y line*=left,
xticklabel style= {rotate=0,anchor=near xticklabel, font=\small},
xticklabel=\year,
xtick distance=365.5,
ytick distance=0.25,
xlabel={Year of CVE assignment},
xlabel style={align=center, font=\small},
ylabel style={align=center, font=\small},
ylabel={\% vulns disc. internal.},
yticklabel={\pgfmathparse{\tick*100}\pgfmathprintnumber{\pgfmathresult}\%}, 
legend style={at={(0,1)},anchor=north west},
]
\addplot[mark=square*, blue, line width=1pt] coordinates {
(2015-01-01, 0)
(2016-01-01, 0.16666666666666666)
(2017-01-01, 0)
(2018-01-01, 0.13333333333333333)
(2019-01-01, 0.13043478260869565)
(2020-01-01, 0.6666666666666666)
(2021-01-01, 0.21052631578947367)
(2022-01-01, 0.25)
(2023-01-01, 0.6046511627906976)
};
\addplot[mark=*, red, line width=1pt] coordinates {
(2018-01-01, 0.911864406779661)
(2019-01-01, 0.7610294117647058)
(2020-01-01, 0.7034883720930233)
(2021-01-01, 0.5074626865671642)
(2022-01-01, 0.5863636363636363)
(2023-01-01, 0.5668789808917197)
};

\addplot[mark=diamond*, green!50!gray, line width=1pt] coordinates {
(2020-01-01, 1)
(2021-01-01, 0)
(2022-01-01, 0.028846153846153848)
(2023-01-01, 0.07096774193548387)
};

\addplot[mark=triangle*, orange, line width=1pt] coordinates {
(2019-01-01, 1)
(2020-01-01, 1)
(2021-01-01, 0.021052631578947368)
(2022-01-01, 0.23018867924528302)
(2023-01-01, 0.10410094637223975)
};

\legend{Samsung, Qualcomm, Unisoc, Mediatek}
\end{axis}
\end{tikzpicture}
\vspace{-1em}
\caption{Vulnerabilities published per year per CM. Bars show the total number of published vulnerabilities, lines show the fraction of vulnerabilities discovered internally by each CM.}
\label{fig:timeseries_discovery}
\end{figure*}
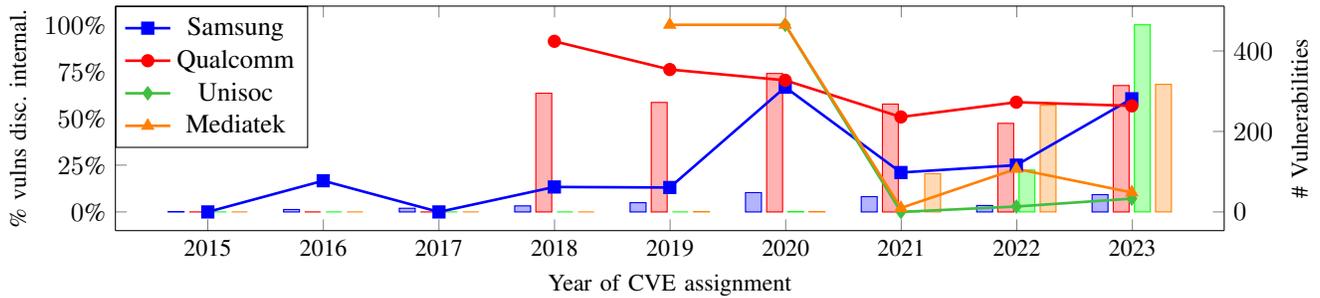

\subsubsection{Growth in vulnerability discovery}\label{sec:who_discovers}

We depict the amount of newly assigned CVEs per chipset manufacturer and year in the bar-chart of Figure~\ref{fig:timeseries_discovery}. Overall, we notice that the amounts of yearly discovered vulnerabilities affecting Qualcomm and Samsung chipsets have stayed constant since 2018, while Mediatek and Unisoc chipsets have seen a significant increase in discovered vulnerabilities. At the same time, we see that Mediatek and Unisoc discover almost no vulnerabilities internally, as illustrated by the line-chart in Figure~\ref{fig:timeseries_discovery}. The growth in absolute numbers of discovered vulnerabilities for both CMs is thus driven by increased scrutiny from external researchers. We see two potential reasons for this increase in external discoveries: First, while Samsung's and Qualcomm's market shares have stayed the same, Mediatek and Unisoc significantly increased their market shares in the period from 2018 to 2023 \cite{statcountermarketshare}, likely attracting more research on their products. Second, while Samsung and Qualcomm have maintained bug bounty programs since 2017 \cite{samsungbugbounty, qualcommbugbounty}, Mediatek and Unisoc vulnerabilities only became eligible for bug bounties in 2023 \cite{mtkbugbounty, googlebugbounty}, potentially further incentivizing external researchers to discover vulnerabilities in their products.

\subsubsection{Internal discovery rates}

In 2023, the products of Mediatek and Qualcomm were both eligible for similar bug bounties \cite{qualcommbugbounty, mtkbugbounty}, and for both CMs around 300 vulnerabilities were published. However, based on the information in CM security bulletins, Qualcomm was able to discover 57\% of these internally, while Mediatek only found 10\% internally. Even more strikingly, Unisoc published 465 vulnerabilities, of which they found only around 7\% internally.
One would expect that the ratio between internal and external discoveries would be roughly equal for all CMs, as long as there are no external factors (\eg bug bounties) that would pull external researchers more towards particular CMs.
There are two potential reasons for the observed discrepancy between CMs: Either, Mediatek and Unisoc indeed almost never find vulnerabilities internally, or they find more vulnerabilities internally than they publicly disclose.
We also observe a positive trend: 
Samsung's ratio of internally discovered vulnerabilities has more than doubled from 25\% in 2022 to 60\% in 2023. 
Although it might be an outlier, as was 2020, this spike could be caused by increased scrutiny following critical vulnerabilities uncovered by Google Project Zero \cite{vergeexynos} and Samsung's establishment of a dedicated chipset product security team in 2023 \cite{samsungpsirt}. %

\begin{table}[t]
    \centering
    \caption{Number of total vulnerabilities grouped by component and manufacturer. Percentage of internally discovered vulnerabilities shown in parenthesis.}%
    \label{tab:component_num_discovered}
    \resizebox{1\columnwidth}{!}{%
    \begin{tabular}{lrrrrr}
    \toprule
        \textbf{Chipset component} & \textbf{Samsung} & \textbf{Qualcomm} & \textbf{Unisoc} & \textbf{Mediatek} & \textbf{Total} \\
     \midrule
Cellular & 54 (57.4\%) & 305 (78.4\%) & 173 (13.9\%) & 70 (20.0\%) & 602 \\
WiFi & 4 (0.0\%) & 356 (59.3\%) & 91 (1.1\%) & 111 (13.5\%) & 562 \\
GPU & 21 (23.8\%) & 254 (58.7\%) & 26 (0.0\%) & 107 (5.6\%) & 408 \\
Trust & 29 (37.9\%) & 186 (79.6\%) & 2 (0.0\%) & 42 (31.0\%) & 259 \\
Audio & 0 (-) & 90 (53.3\%) & 11 (0.0\%) & 19 (47.4\%) & 120 \\
Vision & 2 (0.0\%) & 39 (51.3\%) & 20 (5.0\%) & 40 (10.0\%) & 101 \\
Bluetooth & 3 (66.7\%) & 53 (56.6\%) & 7 (0.0\%) & 22 (13.6\%) & 85 \\
Debug & 3 (0.0\%) & 32 (50.0\%) & 24 (4.2\%) & 15 (0.0\%) & 74 \\
Boot & 12 (8.3\%) & 37 (48.6\%) & 0 (-) & 16 (18.8\%) & 65 \\
IPC & 1 (0.0\%) & 33 (84.8\%) & 0 (-) & 29 (24.1\%) & 63 \\
Machine learning & 10 (30.0\%) & 10 (10.0\%) & 1 (0.0\%) & 41 (2.4\%) & 62 \\
Position & 0 (-) & 18 (61.1\%) & 4 (25.0\%) & 17 (0.0\%) & 39 \\
Memory management & 3 (33.3\%) & 7 (57.1\%) & 3 (0.0\%) & 24 (12.5\%) & 37 \\
Power & 0 (-) & 6 (83.3\%) & 17 (0.0\%) & 11 (36.4\%) & 34 \\
Virtualization & 1 (100.0\%) & 11 (72.7\%) & 0 (-) & 2 (50.0\%) & 14 \\
NFC & 0 (-) & 4 (75.0\%) & 0 (-) & 0 (-) & 4 \\
\midrule
Total & 143 & 1,441 & 379 & 566 & 2,529 \\
    \bottomrule
    \end{tabular}
    }
\end{table}

\subsubsection{Number of discovered vulnerabilities by component}\label{sec:rq2_num_disc_by_component}
To understand if the probability of discovering a new vulnerability differs between a chipset's components, 
we group the number of discovered vulnerabilities by component and manufacturer.
The result is shown in Table~\ref{tab:component_num_discovered}. Most vulnerabilities are discovered in cellular connectivity followed by WiFi and GPUs. This holds true when looking at all manufacturers combined, as well as for every manufacturer independently, except for Samsung. %
Table~\ref{tab:component_num_discovered} also allows us to determine the components in which chipset manufacturers' internal security teams are predominantly discovering vulnerabilities. We observe that, similar to the overall discovery rates, Qualcomm and Samsung surpass the internal discovery rates of Mediatek and Unisoc on almost all components, underlining that their internal discovery rate is universally higher.
\begin{Insight}{Vulnerability discovery}{placeholder2}
\myuline{\textbf{Observations:}}
\begin{compactitem}[$\circ$]
    \item The amount of internally and externally discovered vulnerabilities vary greatly between CMs, with two CMs discovering less than 15\% of reported vulnerabilities internally.\smallskip
    \item There exists a large disparity in number of discovered vulnerabilities between components, with cellular and WiFi vulnerabilities being discovered  most frequently.
\end{compactitem}
\myuline{\textbf{New insights:}}
\begin{compactitem}[$\circ$]
    \item Independent bug bounty hunters, security research companies and academic researchers, rather than in-house product security teams, appear to be the driving force behind vulnerability discovery in the chipsets of some CMs.
    \item The increase in discovered vulnerabilities in recent years is due to external researchers targeting more CMs, rather than improving techniques for a single CM’s chipsets. This suggests that adapting existing techniques to additional CMs’ products is more effective than refining techniques for one CM.
\end{compactitem}
\end{Insight}

\subsection{RQ3: Vulnerability Patching}\label{sec:rq3}

\begin{rqintro}
\noindent\emph{When are patches available and how severe are the chipset vulnerabilities they mitigate?}
\end{rqintro}
\subsubsection{Vulnerability severeness in firmware and drivers}
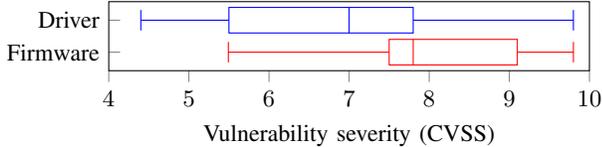
\begin{figure}[t]
\centering
\begin{tikzpicture}
  \begin{axis}
    [
    width=.9\columnwidth,
    height=2.5cm,
    xmin=4,
    xmax=10,
    ytick pos=left,
    xtick pos=bottom,
    xtick align=outside,
    xtick={4,5,6,7,8,9,10},
    xlabel={Vulnerability severity (CVSS)},
    ylabel style={align=center, font=\small},
    xlabel style={align=center, font=\small},
    ytick={1,2},
    yticklabel style={align=right, font=\small},
    xticklabel style={font=\small},
    yticklabels={
    \small{Firmware}, \small{Driver}
    }]
     \addplot+[
      boxplot prepared={
        median=7.8,
        upper quartile=9.1,
        lower quartile=7.5,
        upper whisker=9.8,
        lower whisker=5.4925
      },
      color=red
    ] coordinates {};

    \addplot+[
      boxplot prepared={
        median=7,
        upper quartile=7.8,
        lower quartile=5.5,
        upper whisker=9.8,
        lower whisker=4.4
      },
      color=blue
    ] coordinates {};
  \end{axis}
\end{tikzpicture}
\caption{Severity distribution of firmware and driver vulnerabilities. Severity information is based on NIST analysis.}
\label{fig:boxplot_severity_by_location}
\end{figure}

It has previously been speculated that vulnerabilities in firmware, running on chipset processors outside of the Android OS, are more severe than vulnerabilities in drivers. Prior work argues that this is because drivers, interfacing with chipset processors from inside Android, inherit security features from Android, which firmware does not \cite{weinmann2012baseband}. We now validate whether this claim can be verified empirically by comparing severity scores of driver and firmware vulnerabilities in Figure~\ref{fig:boxplot_severity_by_location}.
While the most severe vulnerabilities of firmware and drivers are capable of enabling a whole system compromise, we observe that the medians of both populations differ, with firmware vulnerabilities showing a median that is 0.8 CVSS points higher than driver vulnerabilities, implying that, on average, firmware vulnerabilities are indeed more severe than driver vulnerabilities. We confirmed this statistically significant difference at $p=0.05$ using a Kruskal-Wallis test.

\subsubsection{Time until patch availability}\label{sec:patch_timeframe}
\begin{figure*}[th]
\centering
\input{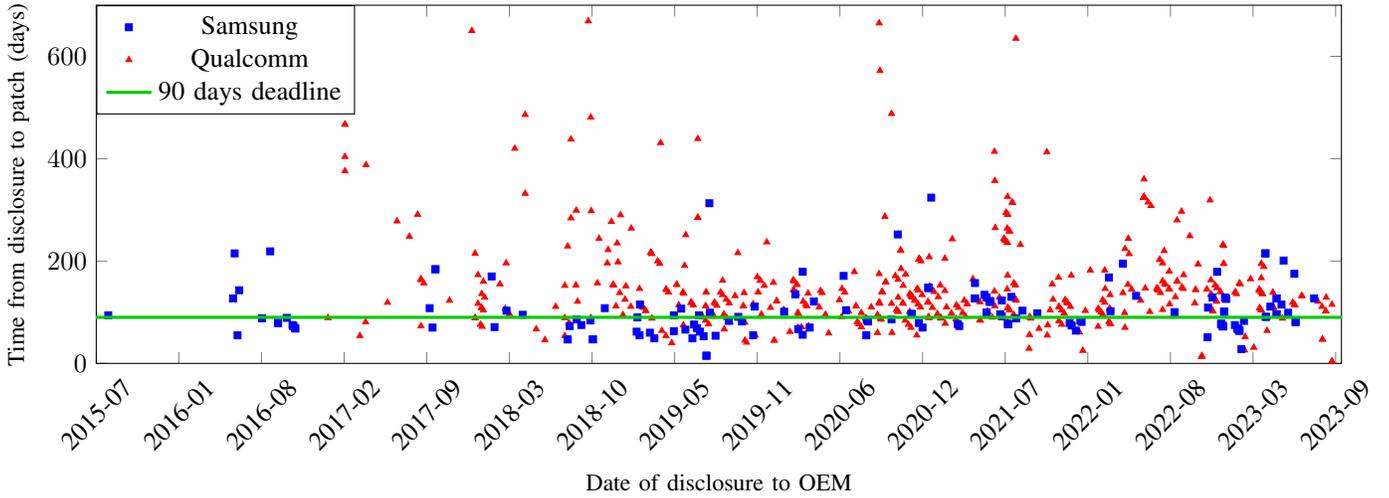}
\vspace{-1em}
\caption{Time frame from vulnerability disclosure to patch availability, each marker representing a single vulnerability. Graph cut-off at $y=700$ days for readability. Mediatek and Unisoc are excluded due to missing discovery date information.}
\label{fig:scatter_notified_to_patch}
\end{figure*}
To verify how long it takes chipset manufacturers to develop a patch and provide it to OEMs, we compare the date $T_\text{report}(v)$ a vulnerability $v$ has been (externally) reported to the date the chipset manufacturer notified the OEM and provided a patch, or they published the vulnerability on their website $T_\text{patch}(v)$. Here we are particularly interested in whether the industry-standard 90-day responsible disclosure period is adhered to, \ie $T_\text{patch}(v)- T_\text{report}(v) \leq 90 \text{ days}$. As evident from Figure~\ref{fig:scatter_notified_to_patch}, neither Qualcomm nor Samsung can consistently meet the 90-day responsible disclosure deadline. Within 90 days, Samsung has patched 46.9\%, while Qualcomm has patched 19.9\% of vulnerabilities reported by external researchers. Instead, Samsung requires
185 days to address 95\% of their vulnerabilities. For Qualcomm, the discrepancy is even more significant, as they require
348 days to supply patches for 95\% of their vulnerabilities. Even more worrisome, we observe that a few remaining outlier vulnerabilities required several years to be addressed. There is no strict correlation between the affected component and the time it takes to develop a patch. Interestingly, our data also shows there is no meaningful correlation between vulnerability severity and patching time frame. This leads us to conclude that high severity vulnerabilities are seemingly not prioritized.

\subsubsection{Information availability and consistency}\label{sec:info_availability}

After a patch has been developed, information on the vulnerability is published by CMs. To verify whether all vulnerability information sources provide the same coherent information, we first compare where vulnerabilities are being published and then assess if the information across all sources is identical. This is important, as inconsistent information on the same vulnerability might lead to confusion and a misjudgement of a vulnerability's severity by OEMs, researchers, and end-users. To do so, we leverage the fact that our knowledge base aggregates information from the NVD, CM's, and the AOSP in directly comparable structure, including publishing dates.

\subsubsection*{Availability} 
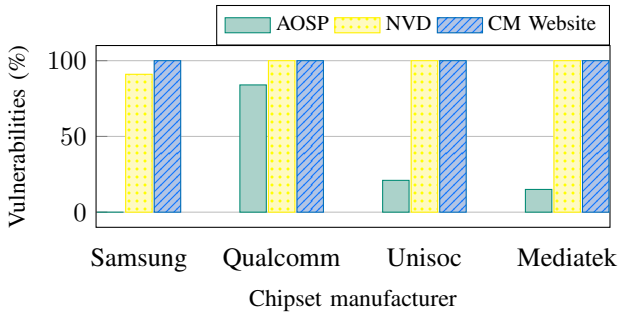
\begin{figure}[th]
\centering
\begin{tikzpicture}
    \begin{axis}[
        width  = 0.95*\columnwidth,
        height = 4cm,
        major x tick style = transparent,
        ybar=2*\pgflinewidth,
        bar width=10pt,
        ymajorgrids = true,
        ylabel = {Vulnerabilities (\%)},
          xlabel style={align=center, font=\small},
  ylabel style={align=center, font=\small},
        xlabel={Chipset manufacturer},
        symbolic x coords={Samsung,Qualcomm,Unisoc,Mediatek},
        xtick = data,
        scaled y ticks = false,
        enlarge x limits=0.1,
        legend cell align=left,
        legend style={
                at={(1,1)},
                anchor=south east,
                legend columns = 3
        }
    ]
        \addplot[area legend, fill=PineGreen!30!white, draw=PineGreen, mark=none]
            coordinates {(Samsung, 0) (Qualcomm,84) (Unisoc,21) (Mediatek,15)};

        \addplot[area legend, preaction={fill, yellow!30!white}, draw=yellow, pattern = dots, pattern color = yellow, mark=none]
             coordinates {(Samsung,91) (Qualcomm,100) (Unisoc,100) (Mediatek,100)};

        \addplot[area legend, preaction={fill, NavyBlue!30!white}, draw=NavyBlue, pattern = north east lines, pattern color = NavyBlue,mark=none]
             coordinates {(Samsung, 100) (Qualcomm,100) (Unisoc,100) (Mediatek,100)};

        \legend{\footnotesize{AOSP}, \footnotesize{NVD}, \footnotesize{CM Website}}
    \end{axis}
\end{tikzpicture}
\vspace{-1em}
\caption{Comparison of vulnerability information availability.}
\label{fig:barchart_data_availability}
\end{figure}

First, we assess if manufacturers reliably publish vulnerabilities on all three platforms.
Our goal is to capture a recent, rather than a historical, picture of the vulnerability publication practises. Over our entire data set, we found that one year is sufficient time to propagate between vantage points for 99.9\% of all vulnerabilities that are eventually included in all vantage points.
We therefore base our report on vulnerabilities published in the time frame from June 2022 to May 2023, such that the vulnerabilities published in May 2023 had until April 2024, \ie one year, to propagate to all vantage points. Vulnerabilities that have not reached all vantage points within this time frame are, according to our data, extremely unlikely  to propagate at all ($<0.1\%$ of cases). The result of this measurement is shown in Figure~\ref{fig:barchart_data_availability}. From this chart, it becomes clear that vulnerabilities published on chipset manufacturer websites are almost always also published in the NVD, and only very few Samsung vulnerabilities are missing from the NVD. Notably, no vulnerabilities are missing from the CM websites, \ie vulnerabilities published in CM security bulletins are a superset of the vulnerabilities published via all other vantage points. However, we notice that Unisoc and Mediatek publish less than 25\% of chipset vulnerabilities in AOSP security bulletins, and Samsung has not published any vulnerability in the AOSP bulletins. The AOSP security bulletins thus are incomplete, and must be augmented with CM (or NVD) website information to gain a holistic picture on which vulnerabilities affect a smartphone.

\subsubsection*{Consistency} As described in Section~\ref{sec:lifecycle}, not only is the severity of a vulnerability analyzed by the chipset manufacturer but also by NIST, which assigns a CVSS score independently. This analysis by NIST has recently been criticized for exaggerating the severity of vulnerabilities in open source projects \cite{edgeboguscve, postgrenota, stenbergnvcmakesup}.
However, out of the 2,249 vulnerabilities for which we obtained a CVSS severeness rating from the CMs' websites as well as from NIST, we observe that for 10\% of these vulnerabilities NIST assigned a lower severity than the CM, and in 15\% of the cases NIST assigned a higher severity. In the realm of chipset vulnerabilities, our data thus does not support the aforementioned criticism that NIST systematically exaggerates severities. 

\begin{Insight}{Vulnerability patching}{insights_rq3}
\myuline{\textbf{Observations:}}
\begin{compactitem}[$\circ$]
    \item On average, vulnerabilities in chipset firmware are more severe than driver vulnerabilities.\smallskip
    \item The 90-day responsible disclosure period is commonly not adhered to by Qualcomm and Samsung.\smallskip
    \item Vulnerability information is often missing from monthly AOSP security bulletins.
\end{compactitem}
\myuline{\textbf{New insights:}}
\begin{compactitem}[$\circ$]
    \item A 90-day disclosure period is inapplicable to vulnerabilities in chipset drivers and firmware, given that it is violated on a regular basis.
    Researchers should not rely on vulnerabilities being addressed within 90 days to assess when to publicly disclose vulnerabilities.
    \item AOSP security bulletins lack comprehensive information on chipset vulnerabilities. This could mislead users and researchers alike to underestimate the amount of vulnerabilities affecting a device.
\end{compactitem}
\end{Insight}

\subsection{RQ4: Vulnerability Updating}\label{sec:rq4}

\begin{rqintro}
\noindent\emph{What are the characteristics of the update process utilized by OEMs to address chipset vulnerabilities in Android devices?}
\end{rqintro}

\subsubsection{Number of affected smartphones}
\begin{figure}[h]
\centering
        \centering
         \begin{tikzpicture}
  \begin{axis}
    [
    width=.9\columnwidth,
    height=3.5cm,
    xmin=0,
    xmax=2222,
    ytick pos=left,
    xtick pos=bottom,
    xtick align=outside,
    xlabel={Number of affected smartphone models},
    ytick={1,2,3,4},
    yticklabel style={align=right},
    yticklabels={
    \small{Qualcomm}, \small{Mediatek}, \small{Samsung}, \small{Unisoc}
    }]
     \addplot+[
      boxplot prepared={
        median=277,
        upper quartile=712,
        lower quartile=86,
        upper whisker=1485,
        lower whisker=0
      },
      color=red
    ] coordinates {};

    \addplot+[
      boxplot prepared={
        median=652,
        upper quartile=1306,
        lower quartile=182,
        upper whisker=2001,
        lower whisker=37
      },
      color=orange
    ] coordinates {};

    \addplot+[
        boxplot prepared={
            median=441,
            upper quartile=441,
            lower quartile=150,
            upper whisker=441,
            lower whisker=6
        },
        color=blue
    ] coordinates {};

    \addplot+[
        boxplot prepared={
            median=152,
            upper quartile=152,
            lower quartile=152,
            upper whisker=152,
            lower whisker=0
        },
        color=green
    ] coordinates {};
  \end{axis}
\end{tikzpicture}
    \vspace{-1em}
    \caption{Number of affected smartphones by a vulnerability, per CM.}
    \label{fig:hist_num_phones_affected}
\end{figure}
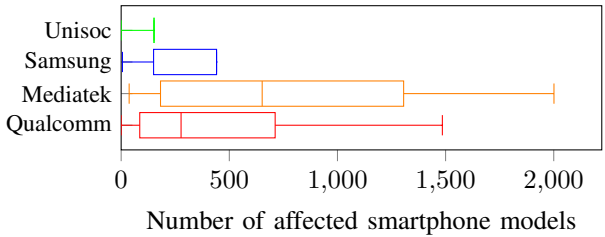

The amount of smartphone models affected by a chipset vulnerability $v$ is determined by two factors: (i) how many chipset models are affected by the vulnerability $|\{c \in C |v \in V(c)\}|$ and (ii) the amount of smartphone models $s \in S$ that contain one of these affected chipset models. Assuming that $B(s) = c$ iff. smartphone model $s$ uses chipset model $c$, then the number of smartphones affected by $v$ is $|\{s \in S | v \in V(B(s))\}| $. We illustrate the resulting number of affected phones per vulnerability and CM in Figure~\ref{fig:hist_num_phones_affected}.
Overall, the median number of affected phones per Mediatek vulnerability (652 phone models) is significantly higher than per Qualcomm vulnerability (277 phone models). Similarly, the most widespread Mediatek vulnerability affected 2,222 smartphone models, while the most prevalent Qualcomm vulnerability affected 1730 smartphone models. In contrast, vulnerabilities affecting Samsung or Unisoc chipsets generally affect the fewer different smartphone models. This is because Samsung predominantly produces chipsets for use in their own smartphones, as well as very few smartphone models made by HTC, Meizu and Motorola. This  limits the amount of potentially affected smartphone models to approximately 450. Likewise Unisoc, being a relatively new CM, only produces chipsets for use in 277 different smartphone models, although supplying to 23 different OEMs.

These results show that a single vulnerability usually has quite drastic ripple effects, as it typically affects multiple chipset models through code re-use (\cf Section~\ref{sec:rq1_origin_of_vulns}), that are used in a variety of smartphone models. The extent of this effect depends on the CM of the chipset.

\subsubsection{Update availability}
CM's may release security advisories containing vulnerability information before OEMs have packaged the patch for the published vulnerabilities into a smartphone update. This would enable threat actors to analyse the, now publicly known, vulnerability and potentially use this information to built exploits, putting the security of end-users at risk.
To quantify this risk, we assess how many vulnerabilities in our dataset seem to have never been mitigated by an update and may still affect devices. All of these vulnerabilities fulfill the following criteria:
\begin{compactitem}
    \item affect at least one smartphone model for which we have update information
    \item are never included in any update changelog, or AOSP security bulletin referenced by an update changelog
    \item published after at least one phone with an affected chipset, for which we have update information, has been released
    \item published before January 2023, to give smartphone manufacturers on a bi-annual update schedule enough time to provide an update
\end{compactitem}

Only 951 (60.2\%) out of 1,546 vulnerabilities matching these criteria have received at least one mitigating update. For the remaining 631 (40.8\%) vulnerabilities, affected users cannot assess whether their devices are still exposed.

\subsubsection{Update timeline}\label{sec:update_timeline}
For the vulnerabilities that have been addressed by updates, we assess for how long end-users have been at risk before an update has been available for their devices.
To this end, for every vulnerability addressed by a device update, we calculate the time from vulnerability announcement on the chipset manufacturer's website until the corresponding update became available. Formally, we define $T_\text{OEM}(v, s)$ as the point in time when an OEM has released an update containing the patch that mitigates vulnerability $v$ in smartphone $s$, and then compute $T_\text{OEM}(v,s) - T_\text{patch}(v)$ for all $(v, s) \in S \times V$ with $v \in V(B(s))$.
We analyzed all 24,226 pairs of vulnerabilities and affected smartphones $(v,s)$ in our knowledge base. The majority of phones will receive an update in under 3 months after a vulnerability is published, as 25\% of affected smartphone models have received an update within 44 days after a vulnerability has been published, and within 71 days 50\% have received updates. However, we observe that updates are sometimes heavily delayed, with the  95\% quantile at 266 days. We also observe that updates for the same vulnerability do not reach all smartphones at the same time. Instead, the median of the interval between the first phone model receiving an update and the last affected phone model receiving an update $\max_{s, s' \in S}T_\text{OEM}(v,s) - T_\text{OEM}(v, s')$ is at 182 days. When we compare the median interval between the first phone model to receive an update, and half of all phones having received an update, we observe a 32 day delay.

Our study thus illustrates two things: First, the update process is not well coordinated between OEMs and CMs, as the CMs publicly announce the existence of a vulnerability before OEMs are able to provide an update to end-users. This allows threat actors to analyze published vulnerabilities, using the time-frame between vulnerability publication and update availability for exploitation. Secondly, it shows that the update process is fragmented, as not all phone models receive an update at once. This could enable threat actors to obtain an update of a phone model that received a timely update and analyze it to gain additional information on exploitability of phone models that are yet-to-receive updates.

\begin{Insight}{Vulnerability updating}{insights_rq4}
\myuline{\textbf{Observations:}}
\begin{compactitem}[$\circ$]
    \item A single vulnerability typically affects hundreds to thousands of different smartphone models.\smallskip
    \item 41\% of vulnerabilities affecting devices in our knowledge base are never addressed in any changelog.\smallskip
    \item Updates reach end-users with a median delay of 71 days after vulnerability publication, with some updates being delayed by more than 8 months.
\end{compactitem}
\myuline{\textbf{New insights:}}
\begin{compactitem}[$\circ$]
    \item End-users are left in the dark regarding a large amount of chipset vulnerabilities, which are seemingly never mitigated via an update. %
    \item Vulnerability announcements by CMs and update availability by OEMs are not well coordinated. This leaves users at risk, when a CM publicly announces a vulnerability, but the OEM takes another 2-8 months to provide an update to end-users.
\end{compactitem}
\end{Insight}

\section{Comparative Analysis}\label{sec:comparison}

We now compare our findings to the results of similar studies in other domains, highlighting key differences, the importance of a better understanding on vulnerability management of chipsets and the gaps that our proposed unified knowledge base fills.

\subsection{Vulnerability Introduction}

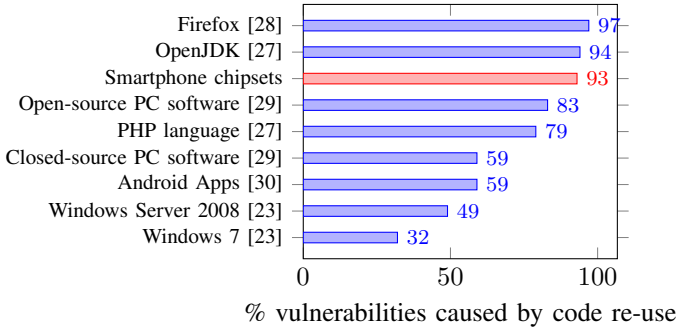
\begin{figure}[h]
\centering
\begin{tikzpicture}
\begin{axis}[ 
xbar, xmin=0,
bar width=4pt,
width=0.65*\columnwidth,
xlabel={\% vulnerabilities caused by code re-use},
ytick={1,2,3,4,5,6,7,8,9},
yticklabels={%
    {Windows 7 \cite{shahzad2019large}},
    {Windows Server 2008 \cite{shahzad2019large}},
    {Android Apps \cite{7802638}},
    {Closed-source PC software \cite{clark2010familiarity}},
    {PHP language\cite{alexopoulos2020tip}},
    {Open-source PC software \cite{clark2010familiarity}},
    {Smartphone chipsets},
    {OpenJDK \cite{alexopoulos2020tip}},
    {Firefox \cite{clark2014moving}}
},
    yticklabel style={align=right, font=\footnotesize},
nodes near coords,
every node near coord/.append style={font=\footnotesize},
nodes near coords align={horizontal},
/pgf/bar shift={0pt},
ytick distance=0.1,
]
\addplot coordinates {
    (32,1)
    (49,2)
    (59,3) 
    (59,4)
    (79,5) 
    (83,6)
    (94,8)
    (97,9)
    };
\addplot coordinates {
(93,7)
};
\end{axis}
\end{tikzpicture} %
\vspace{-1em}
\caption{Prevalence of vulnerabilities inherited from previous software versions through code re-use.}
\label{fig:comparison_rq1}
\end{figure}

In our analysis, we found that code re-use is \emph{the} primary origin of chipset vulnerabilities. As shown in Figure~\ref{fig:comparison_rq1}, code re-use commonly leads to the introduction of vulnerabilities in other ecosystems as well. However, chipset firmware and drivers seem to be particularly prone to this, as their percentage of vulnerabilities from code re-use follows closely those of Firefox \cite{clark2014moving} and OpenJDK \cite{alexopoulos2020tip}. Firefox however follows a rapid release development paradigm, with a new version released every four weeks \cite{firefoxdevmodel}, limiting the amount of potential code changes that can be made in this short time frame. In contrast, chipset development operates on a longer cycle, typically releasing new generations annually with derivative versions appearing every few months. Notably, chipset firmware and drivers are the closed-source products exhibiting the highest proportion of vulnerabilities due to code re-use, compared to Android apps \cite{7802638}, closed-source PC software \cite{clark2010familiarity}, and Microsoft Windows \cite{sivagnanam2021benefits}. \emph{This implies that transitions between versions in closed-source products often result in distinct sets of vulnerabilities for each version. In contrast, chipset firmware and drivers seem to often inherit vulnerabilities, potentially due to chipsets' necessity for backward compatibility with older specifications, such as cellular, Bluetooth, or WiFi standards.} This need for compatibility mirrors that of programming languages like Java (OpenJDK) and PHP, presumably explaining the similarity in vulnerabilities related to code re-use.

\subsection{Vulnerability Discovery}

\subsubsection*{Most affected components} As illustrated in Table~\ref{tab:component_num_discovered}, we found that most vulnerabilities in chipset models are discovered in the cellular modem, followed by WiFi, GPU and trust related functionality. In their study on vulnerabilities in the Android ecosystem, Linares-V{\'a}squez \etal \cite{linares2017empirical} reported that most vulnerabilities found in Android drivers stem from the GPU, WiFi or the camera subsystem, with the cellular modem and trust components never being mentioned. We suspect that this is because the authors obtained their information from the AOSP bulletins which, as we showed in \autoref{sec:info_availability}, lack information on the majority of chipset vulnerabilities. \emph{Prior work therefore severely underestimated the amount of vulnerabilities in various chipset components, such as cellular modems or trust-related components. In contrast, our knowledge base incorporates information from multiple vantage points, enabling us to compile a comprehensive statistic of affected components.}

\subsubsection*{Internal versus external discovery}
In \autoref{sec:who_discovers}, we observed that Mediatek and Unisoc find less than 15\% of vulnerabilities in their products internally. Compared to web browsers, another type of high-risk targets, this rate of internally discovered vulnerabilities appears to be strikingly low. Sivagnanam \etal \cite{sivagnanam2021benefits} measured that 56\% of all Chromium vulnerabilities are found internally by the Chromium development team, Google employees or automated processes. Similarly, Atefi \etal \cite{atefi2023benefits} find that 69\% of all Firefox vulnerability reports come from the Firefox development team internally, rather than external researchers. This shows that, rather than Qualcomm and lately Samsung having an outstandingly well performing internal security team, \emph{Mediatek and Unisoc seem to be under-performing \wrt internally discovered vulnerabilities, compared to other high-impact and high-exposure product categories, such as web browsers}.

\subsubsection*{Impact of bug bounties}
Several studies \cite{alexopoulos2020tip, zhao2015empirical} have shown that in general, the establishment of bug bounty programs leads to an initial increase in discovered vulnerabilities for products covered by the bounty, followed by a decline in discovered vulnerabilities. The authors argue that this is a sign of increasing code quality, as it becomes harder for bug-bounty hunters to discover vulnerabilities. Given Qualcomm's and Samsung's bug bounty programs have existed for 6 to 8 years, one would expect to see a similar decrease in discovered vulnerabilities in their products. However, the amount of vulnerabilities discovered in Qualcomm chipsets is constant, while the quantity of discovered vulnerabilities for Samsung chipsets is following an increasing trend.
\emph{Therefore, it seems that the current bug-bounty programs run by CMs do not significantly reduce the amount of chipset vulnerabilities.} We suspect that this might be due to two factors. First, discovering vulnerabilities in chipsets commonly requires access to hardware for testing purposes (smartphones, software-defined radios \etc), a requirement that does not apply for almost all other software in bug bounty programs and might deter bug hunters. Secondly, chipset vulnerabilities are relatively valuable to parties competing with bug bounty programs for the attention of qualified security researchers, such as exploit brokers. For instance, Zerodium pays up to 200,000 USD for a remote code execution vulnerability in a cellular modem, or up to 100,000 USD for a remote code execution vulnerability in a WiFi component \cite{zerodium}.

\subsection{Patch Development}

In \autoref{sec:patch_timeframe} we found that Qualcomm and Samsung are, on average, able to provide patches for discovered security vulnerabilities within the 90-day responsible disclosure period. This, again, is in stark contrast to the browser ecosystem. On average, the Firefox and Chrome developer teams are able to develop a patch even within 80 days or less \cite{atefi2023benefits}, demonstrating that the 90 days disclosure period is not entirely unrealistic in other ecosystems. Even more strikingly, Zhao \etal \cite{zhao2015empirical} report that 50\% of vulnerabilities reported via bug bounty programs in the web ecosystem are patched within 7 days. In fairness to the CMs, there are several additional steps in the patch development process of chipsets, compared to browsers or web applications: Web browsers or websites do not have do undergo integration tests for hundreds of different phone models, as they typically only target a handful of operating systems. They also do not have to potentially undergo the same regulatory (re-)certification for radio emitters and protocol compliance that chipsets do. \emph{Nonetheless, this highlights that patching chipset vulnerabilities takes significantly longer than patching other high-value targets.}

\subsection{Updates}

\subsubsection*{Web browsers and operating systems}
As illustrated in Section~\ref{sec:update_timeline}, OEMs require 52 days to supply 50\% of affected smartphone models with an update after a vulnerability has been published by a CM. This sharply differs from other types of software. Shazat \etal \cite{shahzad2019large} found that 96\% of Chrome and 58\% of Firefox vulnerabilities receive a mitigating update on the day they are published, while Microsoft and Apple are able to immediately provide updates for their operating systems for 76\%-78\% of the vulnerabilities they publish. We conclude that the additional 52 days required by OEMs are due to the supply chain from CM to OEM that does not exist for web browsers or operating systems, where the party developing a patch is also packaging the patch into an update.

\subsubsection*{Android updates}
When comparing our findings to prior work on Android device updates, we observe that prior work has been underestimating the amount of vulnerabilities impacting a smartphone. Acar \etal \cite{50shades} have highlighted, as part of a case study, that the Qualcomm SM8350 used in the Samsung Galaxy Z Fold3 5G phone suffers from 11 chipset vulnerabilities. However, in their study, the NVD CVE database was used as the sole source, where only one of two possible CPE identifiers to filter for CVEs affecting the SM8350 was employed. Our unified knowledge base reveals that this chipset is actually affected by, at least, 131 vulnerabilities, an increase by more than a factor of 11. This highlights the relevance of cross-validating information from several vantage points, thus validating our proposal for the need to consolidate all the information in one single place. Prior work on Qualcomm chipset vulnerabilities \cite{farhang2020empirical} also found that, while some Qualcomm chipset vulnerabilities are never resolved in specific phones, every CVE is at least addressed by one OEM. In contrast, we observe that for 43\% of CVEs no OEM publishes an update, and thus highlight that the situation has been previously underestimated. A large contributing factor to this underestimation is that Samsung, Unisoc and Mediatek - absent from studies in prior work - only report very few of their vulnerabilities in AOSP security bulletins. However, in the updating phase of the vulnerability lifecycle OEMs rely heavily on these AOSP bulletins to communicate which vulnerabilities have been addressed in their devices, and thus \emph{the absence of chipset vulnerabilities in AOSP bulletins results in missing information for end-users.}

\section{Discussion}\label{sec:discussion}

\subsection{Changes in the Android ecosystem}

We observe two primary ways in which chipsets directly impacted security-relevant changes in the AOSP within the last years.
First, the Android version developed by the AOSP is heavily customized by the CMs and OEMs to run on the devices' chipset. Prior to Android 8, every Android update had to be customized for each phone model, such that it properly runs on the respective phone model's chipset.
As our results show, resolving 95\% of all chipset-related security vulnerability takes OEMs and CMs between 180 days and a full year. To reduce the dependence of Android system updates on CMs, the APIs that are used within the chipset-specific customizations have been gradually standardized since Android 8 \cite{yim2019treble}. This decoupling allows OEMs to deliver Android updates without waiting for CMs to adapt the new version to the chipset, accelerating deployment of general Android system updates.

Secondly, AOSP offers to include vulnerabilities affecting chipsets into the Android Security Bulletins to mitigate some of the heterogeneity and offer users a CM and OEM independent way of assessing the patch level of their device \cite{aospbulletins}. According to our knowledge base, 35\% of all chipset vulnerabilities were included in AOSP security bulletins in 2019 and the inclusion ratio rose to 59\% in 2021, but has since declined to 25\% in 2023. This drop is caused by increasing number of discovered vulnerabilities affecting Mediatek and Unisoc, who do not publish all their vulnerabilities in AOSP bulletins.
As we have seen, vulnerabilities included into AOSP bulletins are less likely to be never mitigated. In this sense, the AOSP bulletins are an organizational measure to improve transparency, but also positively correlate with update availability in cases where CMs commit to publishing all vulnerabilities within these bulletins.

\subsection{Recommendations to Industry}\label{sec:suggestions}

\noindent\textbf{Bolster CMs' internal security teams.} Our study shows that vulnerabilities exist in several chipset generations before discovery and remain in the firmware and drivers until identified (\cf Section~\ref{sec:rq1}). There is an urgent need for increased efforts and resources for faster vulnerability discovery. Currently, half of the CMs discover less than 15\% of vulnerabilities internally, relying on external researchers (\cf Section~\ref{sec:rq3}). This contrasts with the security practices of other high-impact targets like web browsers (\cf \autoref{sec:comparison}). We recommend that CMs with low internal discovery rates conduct internal security testing, such as employing a ``red-team'' of penetration testers, and publish all internally found vulnerabilities in their security bulletins.

\noindent\textbf{Prioritize chipset firmware.} Vulnerabilities in firmware are more severe than driver vulnerabilities, putting devices at greater risk (\cf \autoref{sec:rq3}). We recommend that CMs deploy appropriate technical measures to reduce the impact of firmware vulnerabilities. Potential technical measures that could reduce the severity gap between drivers and firmware include defense techniques such as process isolation and memory-tagging for address sanitization. These are already deployed in Android itself \cite{androidhardening} and thus promote driver security, but are often missing in chipset firmware \cite{hernandez2022firmwire}.

\noindent\textbf{Ensure completeness of AOSP security bulletins.} Information on previously discovered vulnerabilities, patches, and updates is often unavailable or scattered across multiple sources. We find the incomplete information in AOSP security bulletins particularly problematic, as they are the premier way for Android users and researchers to determine if a device has received all the latest security updates. If particular vulnerabilities never occur in these bulletins, it is impossible for users to judge the security of their device. Per our analysis, this is the case for more than 75\% of vulnerabilities affecting chipsets of some CMs (\cf Sections~\ref{sec:knowledge_base}, \ref{sec:rq3} and \ref{sec:rq4}). Completeness of AOSP bulletins could be achieved by enforcing that CMs report all vulnerabilities %
to the AOSP by making this a requirement for Android compatibility certifications.

\noindent\textbf{Establish an industry-wide responsible disclosure timeframe.} While the 90-day responsible disclosure period might be insufficient for hardware deployed in hundreds of different device models, we suggest that an appropriate timeframe (around 200 days, as per our analysis) should be established and followed to minimize outlier vulnerabilities that remain unpatched for extended periods. Ideally, this embargo time-frame would not only include the patch development, but also the update packaging phase of the vulnerability lifecycle, as we have demonstrated in our analysis that this phase adds another significant delay until updates finally arrive on end-users' devices (\cf Section~\ref{sec:rq4}), putting them at risk of threat actors developing exploits based on publicly available vulnerability information.

\subsection{Use Cases in Research}\label{sec:use-cases}

\noindent Our data enables multiple practical use cases that we believe to benefit the research community.

\noindent\textbf{More representative evaluations.} Evaluating novel vulnerability discovery techniques warrants a representative set of devices to empirically test the success of said techniques. As we observed in \autoref{sec:rq3}, many chipset models share the same vulnerabilities through code re-use. Manually testing chipsets affected by mostly overlapping sets of vulnerabilities is time consuming, unnecessarily expensive and thus inefficient. Our website offers a tool (\url{www.chipsets.org/devices/pick}) to effortlessly, and via a graphical user interface, select a variety of devices with chipsets that share fewer vulnerabilities, increasing the likelihood of testing novel implementations rather than re-used ones.
\vspace{1.5mm}

\noindent\textbf{Accurate depiction of the impact of newly discovered vulnerabilities.} Researching chipset vulnerabilities is a dynamic field, and understanding their real-world impact is crucial. Typically, researchers gauge this impact by assessing how many devices are affected by a vulnerability they uncover. However, CMs usually only disclose information on affected chipset models, not the specific smartphone models impacted. Consequently, many research papers on novel vulnerabilities tend to underestimate the number of affected smartphone models due to the cascading effects discussed in \autoref{sec:rq4}, and only report the devices they manually tested. In contrast, our unified knowledge base allows for automatic cross-referencing of chipset vulnerability information published by CMs with affected smartphone models. This provides a comprehensive number of affected device models. We summarize the difference between several papers' reported and actual number of affected devices in Table~\ref{tab:use-case-1}, further highlighting the extent of this under-estimation in current literature. 
\vspace{1.5mm}

\begin{table}[t]
    \centering
    \caption{Presented and actual number of smartphone models affected by chipset CVEs discovered by academia. $\dagger$ identifier as per original paper.}%
    \label{tab:use-case-1}
    \resizebox{1\columnwidth}{!}{%
    \begin{tabular}{llrrrr}
    \toprule
        \textbf{Paper} & \textbf{Identifier} & \textbf{\# aff. (paper)} & \textbf{\# aff. (actual)}\\
     \midrule
 Braktooth \cite{garbelini2022braktooth} & CVE-2021-30348 & 1 & 293\\
 Braktooth \cite{garbelini2022braktooth} & CVE-2022-20021 & - & 609\\
 Instructions Unclear \cite{klischies2023instructions} & CVE-2022-26446 & 2 & 1492 \\
 Instructions Unclear \cite{klischies2023instructions} & CVE-2022-32591 & 1 & 1397 \\
Owfuzz \cite{cao2023owfuzz} & CVE-2021-1903 & 3 & 572 \\
 HW-backed Heist \cite{ryan2019hardware} & CVE-2018-11976 & 1 & 1481 \\
 DoLTEst \cite{park2022doltest} & CVE-2019-2289 & 17 & 1381 \\
 Signal Overshadowing \cite{236354} & SigOver $\dagger$ & 10 & 330\\
 DIKEUE \cite{hussain2021noncompliance} & E1 $\dagger$ & 7 & 420 \\
 DIKEUE \cite{hussain2021noncompliance} & E13 $\dagger$ & 3 & 115 \\
    \bottomrule
    \end{tabular}
    }
\end{table} 

\noindent\textbf{Identifying avenues for future research.} Since our unified knowledge base helps to paint a complete picture of chipset vulnerabilities, this information can therefore further facilitate the identification of outliers that warrant future research. \emph{Underrepresented components:} As described in \autoref{sec:rq3}, so far, extremely few vulnerabilities in the Near Field Communication (NFC) stack have been discovered, although all of these were rather severe. Therefore, NFC might call for further scrutiny. \emph{High severity targets:} Similarly, we found that firmware vulnerabilities tend to be more severe than driver vulnerabilities. However, currently, our data set contains roughly twice as many driver than firmware vulnerabilities. %
We thus identify a need for more approaches on firmware security testing.

\subsection{Threats to Validity}

\noindent\textbf{Undiscovered vulnerabilities.}
As we exclusively analyze vulnerabilities acknowledged by CMs or OEMs, this precludes the estimation of total vulnerabilities in chipsets or phones due to undiscovered vulnerabilities.
\myuline{Potential impact:} For this reason, any absolute number associated with vulnerability introduction (RQ1) does not represent the overall number of existing vulnerabilities. We believe our conclusion is valid nonetheless because the relative numbers in \autoref{sec:rq1} and the significant differences in vulnerability origin and persistence are unaffected, as both discovered and undiscovered vulnerabilities come from the same underlying distribution of chipset vulnerabilities.

\noindent\textbf{Zero-day vulnerabilities.}
Our data set does not contain information on vulnerabilities that were discovered by threat-actors, and that CMs are currently unaware of. Such Zero-day vulnerabilities might be particularly severe vulnerabilities, as threat-actors are typically interested in exploitability.
\myuline{Potential impact:} Within RQ2 we only consider vulnerabilities discovered by CMs (internal) and external researchers who report their findings to the CMs. Therefore, our results do not allow claims on the total amount of discovered vulnerabilities. 
Likewise, if there would be a significant amount of high-severity zero-day vulnerabilities, the absolute numbers of high-severity vulnerabilities within RQ3 and especially \autoref{fig:boxplot_severity_by_location} might be lower-bounds. Nonetheless the conclusion that firmware vulnerabilities are, on average, more severe than driver vulnerabilities prevails, as the underestimation would equally affect firmware and drivers.

\noindent\textbf{Silent patching.}
Our knowledge base does not contain information on vulnerabilities that are known to CMs, but have not been publicly disclosed and instead were patched without a public announcement. %
Given that Qualcomm and Samsung already published hundreds of internally discovered vulnerabilities, some of which being very severe and exploitable, the upside of additionally engaging in silent patching seems to be rather minimal for these two CMs. %
\myuline{Potential impact:} We assume that silently patched vulnerabilities would be internally discovered, as preventing external researchers from publishing information on their discoveries is not generally possible. This would skew the results presented in \autoref{sec:rq2}. Likewise, the results on timeframes for patch development (\autoref{sec:rq3}) and update availability (\autoref{sec:rq4}) may not apply for silently patched vulnerabilities. Lastly, CMs may be particularly inclined to silently patch high severity vulnerabilities. The absolute numbers presented in \autoref{fig:boxplot_severity_by_location} likely represent a lower bound, yet this does not alter the conclusion that firmware vulnerabilities are generally more severe than driver vulnerabilities, as both are equally impacted by this limitation.

\noindent\textbf{Data collection and augmentation.}
Some our our data is obtained from websites, requiring us to automatically extract relevant information.
Since this process is based on carefully written, individual parsers and validators for each vantage point, we believe our results to be as reliable as state of the art papers on this topic, that all employ a similar data collection method \cite{50shades, farhang2020empirical, hou2023can, linares2017empirical}. \myuline{Potential impact:} As information on vulnerabilities is gathered from several independent vantage points (CM security bulletins, AOSP security bulletins and NVD data), we believe that this information is comprehensive. Information that is collected from a single source, such as chipset release dates and assignments of chipset models to phone models, may occasionally be incorrect. Augmented information may also have inaccuracies. However, given the large quantities of vulnerabilities,  chipsets, and phones suggested, the primary outcomes of this study will not be affected by a limited number of data imperfections.

\section{Conclusion}

We create the first comprehensive knowledge base of 3,676 smartphone chipset vulnerabilities, tightly linking chipset models and vulnerabilities to devices and their updates. By empirically analyzing this data set, we discover that chipset vulnerabilities often persist across many chipset generations, such that each chipset vulnerability exposes hundreds or even thousands of different smartphone models. 
We identified significant shortcomings in current vulnerability management processes, such as inconsistently published information and OEMs frequently failing to inform users about vulnerabilities affecting their devices, leaving them at risk of exploitation. Our findings highlight unique characteristics of chipset vulnerabilities, including extensive code reuse and prolonged patch propagation times, which set them apart from vulnerabilities in other high-risk targets like web browsers and operating systems. 
To prevent chipset firmware and drivers from remaining a critical weakness in the security architecture of Android smartphones, we propose several measures to enhance their security.
We envision that our unified knowledge base will enable future research to tackle these problems by identifying particularly problematic aspects of the vulnerability lifecycle.

\section*{Acknowledgment}
We would like to thank our anonymous reviewers for their
valuable comments and suggestions. We also thank Paul Staat, Alyssa Milburn and Marius Muench for their feedback. This work was supported by the German Federal Office for Information Security (FKZ: Pentest-5GSec - 01MO23025B) and the Deutsche Forschungsgemeinschaft (DFG, German Research Foundation) under Germany's Excellence Strategy - EXC 2092 CASA - 390781972.

\bibliographystyle{IEEEtran}
\bibliography{references}

\appendix
In this section we explain further technical details on the implementation that automatically performs the data collection process described in \autoref{sec:method}.

\subsection{Data collection}\label{appendix:jsonhtml}

As shown in \autoref{tab:vantage_point_urls}, we incorporate information from CMs, OEMs, the AOSP, NIST, as well as websites listing device and chipset information.  The entire data collection is implemented in NodeJS\footnote{\url{https://nodejs.org/}}, using the NestJS\footnote{\url{https://nestjs.com/}} framework, using a MariaDB\footnote{\url{https://mariadb.org/}} instance to store collected information using a relational data model. As the websites differ in data formats, we employ different techniques to obtain structured information.
In the cases of Tecno and the NIST NVD, vantage points are available to us as easily processable JSON documents, which we download using NodeJS built-in functionality. The other vantage points require processing of HTML-based websites. All vantage points that are HTML-based websites use a reoccurring HTML structure within the same vantage point. We download their HTML files, again using built-in functionality, and then use the Cheerio library\footnote{\url{https://cheerio.js.org/}} to extract specific fields from the HTML structure via their CSS selector.

\begin{figure}[!ht]
\centering
\includegraphics[width=0.95\columnwidth]{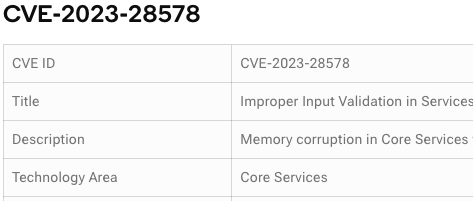}
\caption{Excerpt from a Qualcomm security bulletin. Source: \url{https://docs.qualcomm.com/product/publicresources/securitybulletin/march-2024-bulletin.html}.}
\label{fig:qualcomm_example}
\end{figure}

For instance, all Qualcomm security bulletins use the same, tabular format. This enables us to obtain their information by traversing the HTML structure using CSS selectors, taking account labels - such as HTML column headers, to detect structural changes. \autoref{fig:qualcomm_example} depicts a part of a Qualcomm security bulletin. To obtain the text associated with the Technology Area field, we use the \texttt{td:contains(Technology Area) + td} CSS selector, which selects the table cell adjacent to the cell that contains the text ``Technology Area''. 

For each vantage point we developed a separate parser, taking into account the vantage point's specific structure. This is in-line with the data collection methods employed in prior work \cite{50shades, farhang2020empirical, hou2023can, linares2017empirical}. Any data point obtained is automatically checked for plausibility, allowing us to catch changes should the HTML structure change in the future. For instance, CVE numbers should always follow the \texttt{CVE-<YEAR>-<NUM>} pattern, where the first set of digits \texttt{<YEAR>} must be a valid year in the past.

In cases where the security bulletins are not available as static HTML, and instead are rendered via Javascript, we use Puppeteer\footnote{\url{https://pptr.dev/}}. This allows us to automate a browser which is running in the background to open the relevant website, execute any Javascript to generate HTML, and then process the HTML as described before. This is currently only needed to process security bulletins published by Unisoc.
In contrast to prior work, all collection procedures, are performed periodically, and followed by a transformation into the schema illustrated in \autoref{fig:method_information_relations}.
The unified knowledge base is therefore constantly evolving as vantage points are updated by NIST, CMs, OEMs and the AOSP, rather than being a snapshot of the situation at the time the paper was written.

\subsection{Data augmentation}\label{appendix:location_component}

After the individual data points have been collected, some information is missing or inconsistently represented across CMs. Below, we provide a list of all data fields that we augment to mitigate this issue, and details on the augmentation procedures.

\emph{Chipset components.} When comparing vulnerabilities across CMs, we have to normalize the name of the component affected by a chipset, as different CMs use different names for components implementing the same functionality. For instance, while MediaTek licenses the ARM Mali GPU architecture, Qualcomm uses their in-house Adreno GPU design. As both fulfil the same functional objective, we want them to be assigned a common component name. To fulfil this requirement, our list of components is inherently based on the differentiation used by CMs. We build a list of common component names, and CM specific key-terms identifying components, by iteratively scanning CM security bulletins of all CMs for the field containing the CM specific component name, e.g. "Technology Area" in Qualcomm Security Bulletins. We then perform the following steps:
\begin{compactenum}
    \item Check if the CM specific component name is already in our set of key-terms. If so, stop the process for the current bulletin and continue with the next bulletin.
    \item If not, look up the CM specific component name in technical documentation, source code fragments and marketing material to identify its functionality.
    \item Identify if there is already a common component name for the identified functionality, if not: add it to the list.
    \item Add the CM specific component name as a key term for the common component name either identified or added in step 3.
    \item Continue this process with the next bulletin.
\end{compactenum}

Using this approach, we establish a nomenclature of Android smartphone chipset components that differentiates between the following components:

\noindent\textbf{Bluetooth.} Bluetooth connections and packets, including Bluetooth low energy.

\noindent\textbf{WiFi.} IEEE-802.11 (WiFi) connectivity

\noindent\textbf{Cellular.} Cellular connectivity subsystem, including signal processing, baseband processor and Radio Interface Layer (RIL), used to establish connections to \eg 4G and 5G networks.

\noindent\textbf{GPU.} Subsystem responsible for graphics processing and rendering, including Digital Rights Management (DRM) for visual content (\ie WideVine).

\noindent\textbf{Vision.} Camera, facial recognition and similar visual perception systems.

\noindent\textbf{NFC.} Near-Field-Communication, for instance used for digital payment and smartcard applications.

\noindent\textbf{Boot.} Bootloader and other early-stage firmware used to start the operating system.

\noindent\textbf{Position.} Subsystem to determine device position, \eg via the Global Positioning System.

\noindent\textbf{Audio.} Input and output audio processing pipelines.

\noindent\textbf{Virtualization.} Hardware support for virtual machine monitors and hypervisors.

\noindent\textbf{Machine Learning}. Support for the implementation of inference systems,\eg acceleration of matrix operations for neural network computations via dedicated hardware.

\noindent\textbf{Trust.} Isolation functionalities for programs with elavated security requirements (\ie via ARM TrustZone), signature as well as efuse mechanisms, used for instance to authenticate device integrity.

\noindent\textbf{Power.} Power management facilities, for instance used to reduce power consumption while the device is not actively used.

\noindent\textbf{IPC.} Inter-processor communication, used by primary components to communicate with each other, e.g. via mailbox mechanisms or queues transmitted via serial busses.

\noindent\textbf{Memory.} Memory management and protection mechanisms.

\noindent\textbf{Debug.} Logging and debugging interfaces.

\subsection{Data integration}\label{appendix:integration}

\begin{figure*}[!ht]
\centering
\includegraphics[width=0.95\textwidth]{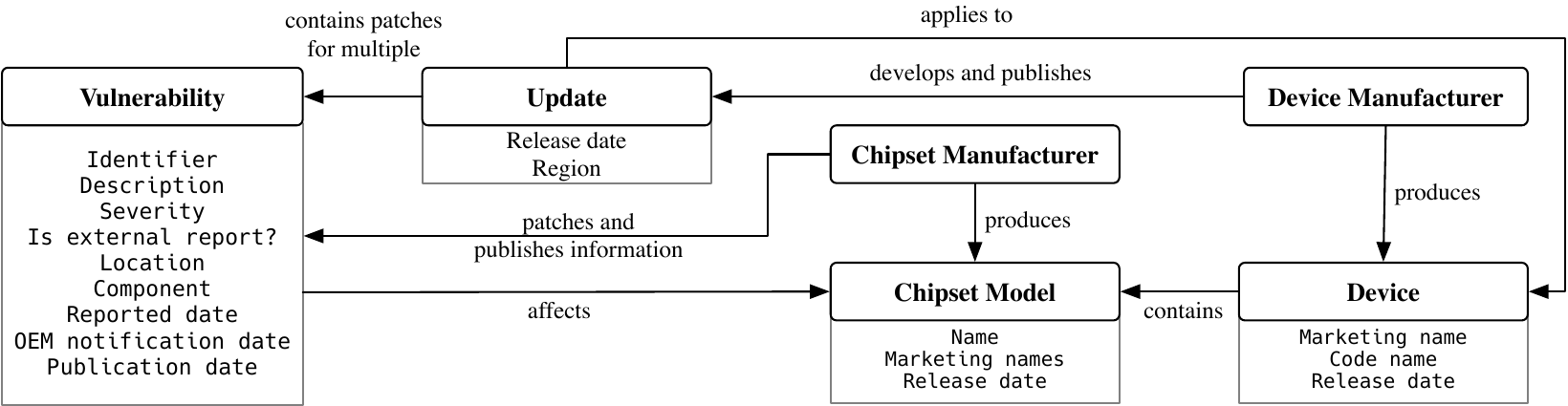}
\caption{Relational schema employed by our knowledge base. Boxes correspond to relations, and contain the attributes in that relation. Arrows correspond to foreign key references. }
\label{fig:method_information_relations}
\end{figure*}

After the data has been collected and augmented, information collected from different vantage points is still separated, and not adhering to a common schema. During the data integration step, we integrate individually collected data points into the shared relational schema shown in \autoref{fig:method_information_relations}. To do so, we first map the individual descriptors used by the vantage points to our own, canonical descriptors. For instance, Qualcomm's \texttt{CVE ID} in \autoref{fig:qualcomm_example} becomes \texttt{Identifier} in the \texttt{Vulnerability} relation.
The mapping between canonical and vantage point-specific attribute names is based on a set of static rules. We then insert the collected information into the relational database. For potentially conflicting information, such as vulnerability severity, we store the values separately for each vantage point. This allows us to analyze conflicting information in \autoref{sec:rq3}.
Lastly, we establish foreign key relationships between the different entities, which allows us to analyze the vulnerability lifecycle based on data from different vantage points in a unified way. For instance, we establish a many-to-many relationship between vulnerabilities and the chipset models they affect. To do so, we use the list of affected chipset model names published in the CM security bulletin and NVD entry of each vulnerability to identify matching chipset models in our database. We then establish the foreign key relationship between the vulnerability, and all identified chipset models. Since we also establish a one-to-to-many relationship between every chipset model and the devices it is used in, this allows us to determine which vulnerability affects which devices by traversing the foreign key relationships. All foreign keys we establish are depicted as arrows in \autoref{fig:method_information_relations}.

\begin{table*}[!ht]
    \centering
    \caption{List of vantage points}%
    \label{tab:vantage_point_urls}
    \begin{tabular}{lllcccc}
        & & &
         \multicolumn{4}{c|}{\cellcolor{black}\color{white}\textbf{Relevant for Phase}}
         \vspace{0.1em}
        \\
        \textbf{Name} &
        \textbf{Vantage Points} &
        \textbf{Format} &
        \textbf{Intro.} &
        \textbf{Discovery} &
        \textbf{Patching} &
        \textbf{Updating}
        \\
        \midrule
        Qualcomm Security Bulletin\footnotemark[1] &
        CM &
        HTML &
        \cmark &
        \cmark &
        \cmark &
        \xmark
        \\
        Mediatek Security Bulletin\footnotemark[2] &
        CM &
        HTML &
        \cmark &
        \cmark &
        \cmark &
        \xmark
        \\
        Samsung Security Bulletin\footnotemark[3] &
        CM, OEM &
        HTML &
        \cmark &
        \cmark &
        \cmark &
        \cmark
        \\
        Samsung Semicon. Security Bulletin\footnotemark[4] &
        CM &
        HTML &
        \cmark &
        \cmark &
        \cmark &
        \xmark
        \\
        Unisoc Security Bulletin\footnotemark[5] &
        CM &
        Javascript &
        \cmark &
        \cmark &
        \cmark &
        \xmark
        \\
    \midrule
    Samsung Updates\footnotemark[6] & OEM & HTML & 
    \xmark & \xmark & \xmark & \cmark \\
    Xiaomi Updates\footnotemark[7] & OEM & HTML & \xmark & \xmark & \xmark & \cmark \\
    Tecno Updates\footnotemark[8] & OEM & JSON & \xmark & \xmark & \xmark & \cmark \\
    Tecno Changesets\footnotemark[9] & OEM & JSON &  \xmark & \xmark & \xmark & \cmark \\

     \midrule
        Android Security Bulletin\footnotemark[10] &
        AOSP &
        HTML &
        \xmark &
        \xmark &
        \cmark &
        \cmark \\
    \midrule
        NIST NVD Database\footnotemark[11] &
        NIST, CVE &
        JSON &
        \cmark &
        \cmark &
        \cmark &
        \xmark \\  

    \midrule
        GSMArena\footnotemark[12] &
        Device information &
        HTML &
        \xmark &
        \xmark &
        \xmark &
        \cmark \\  

        Wikipedia\footnotemark[13] &
        Chipset release dates &
        HTML &
        \cmark &
        \xmark &
        \xmark &
        \xmark \\
     
    \midrule
    \textbf{Exemplary URLs:} \\
    \multicolumn{6}{l}{\footnotemark[1] \url{https://docs.qualcomm.com/product/publicresources/securitybulletin/march-2024-bulletin.html}} \\
    \multicolumn{6}{l}{\footnotemark[2] \url{https://corp.mediatek.com/product-security-bulletin/March-2024}}
    \\
    \multicolumn{6}{l}{\footnotemark[3] \url{https://security.samsungmobile.com/securityUpdate.smsb}}
    \\
    \multicolumn{6}{l}{\footnotemark[4] \url{https://semiconductor.samsung.com/support/quality-support/product-security-updates/}} \\
    \multicolumn{6}{l}{\footnotemark[5] \url{https://www.unisoc.com/en_us/secy/announcementDetail/1754320321801945089}} \\
    \multicolumn{6}{l}{\footnotemark[6] \url{https://doc.samsungmobile.com/SM-A415F/PHN/doc.html}} \\
    \multicolumn{6}{l}{\footnotemark[7] \url{https://xiaomifirmwareupdater.com/archive/miui/agate/}} \\
    \multicolumn{6}{l}{\footnotemark[8] \url{https://security.tecno.com/slm/deviceScope?lang=en-US}}\\
    \multicolumn{6}{l}{\footnotemark[9] \url{https://security.tecno.com/slm/patchInfo?lang=en_US&year=2022&quarter=01}}\\
    \multicolumn{6}{l}{\footnotemark[10] \url{https://source.android.com/docs/security/bulletin/2023-09-01}}\\
    \multicolumn{6}{l}{\footnotemark[11] \url{https://services.nvd.nist.gov/rest/json/cves/2.0?cveId=CVE-2022-33251}}\\
    \multicolumn{6}{l}{\footnotemark[12] \url{https://www.gsmarena.com/samsung_galaxy_s20_fe_5g-10377.php}}\\
    \multicolumn{6}{l}{\footnotemark[13] \footnotesize{\url{https://en.wikipedia.org/w/api.php?action=parse&page=List_of_Qualcomm_Snapdragon_systems_on_chips&prop=text&section=1&format=json}}}\\

    \end{tabular}
\end{table*}

\end{document}